\documentclass[twocolumn]{aastex63}
\usepackage{multirow}
\usepackage{lineno}
\usepackage[caption=false]{subfig}
\usepackage{amsmath}

\shorttitle{COSI Detection of Galactic Positron Annihilation}
\shortauthors{Kierans et al.}

\begin{document}

\title{Detection of the 511 keV Galactic positron annihilation line with COSI}

\correspondingauthor{Carolyn A. Kierans}
\email{carolyn.a.kierans@nasa.gov}

\author{C. A. Kierans}
\affil{NASA/Goddard Space Flight Center, Greenbelt, USA}
\author{S. E. Boggs}
\affil{University of California, San Diego, USA}
\author{A. Zoglauer}
\affil{Space Sciences Laboratory, University of California, Berkeley, USA}
\author{A. W. Lowell}
\affil{University of California, San Diego, USA}
\author{C. Sleator}
\affil{Space Sciences Laboratory, University of California, Berkeley, USA}
\author{J. Beechert}
\affil{Space Sciences Laboratory, University of California, Berkeley, USA}
\author{T. J. Brandt} 
\affil{NASA/Goddard Space Flight Center, Greenbelt, USA}
\author{P. Jean}
\affil{IRAP Toulouse, France}
\author{H. Lazar}
\affil{Space Sciences Laboratory, University of California, Berkeley, USA}
\author{J. Roberts}
\affil{University of California, San Diego, USA}
\author{T. Siegert}
\affil{University of California, San Diego, USA}
\author{J. A. Tomsick}
\affil{Space Sciences Laboratory, University of California, Berkeley, USA}
\author{P. von Ballmoos}
\affil{IRAP Toulouse, France}

\begin{abstract}


The signature of positron annihilation, namely the 511 keV $\gamma$-ray line, was first detected coming from the direction of the Galactic center in the 1970's, but the source of Galactic positrons still remains a puzzle. 
The measured flux of the annihilation corresponds to an intense steady source of positron production, with an annihilation rate on the order of $\sim10^{43}$~e$^{+}$/s.
The 511 keV emission is the strongest persistent Galactic $\gamma$-ray line signal and it shows a concentration towards the Galactic center region. An additional low-surface brightness component is aligned with the Galactic disk; however, the morphology of the latter is not well constrained.
The Compton Spectrometer and Imager (COSI) is a balloon-borne soft $\gamma$-ray (0.2--5 MeV) telescope designed to perform wide-field imaging and high-resolution spectroscopy. One of its major goals is to further our understanding of Galactic positrons. 
COSI had a 46-day balloon flight in May--July 2016 from Wanaka, New Zealand, and here we report on the detection and spectral and spatial analyses of the 511 keV emission from those observations. 
To isolate the Galactic positron annihilation emission from instrumental background, we have developed a technique to separate celestial signals utilizing the COMPTEL Data Space.
With this method, we find a 7.2$\sigma$ detection of the 511 keV line. We find that the spatial distribution is not consistent with a single point source, and it appears to be broader than what has been previously reported.

\end{abstract}



\section{Introduction} 
\label{sec:intro}

The 511~keV signature of electron-positron annihilation was first observed from the Galactic center (GC) region in the 1970's \citep{johnson1972, haymes1975, leventhal1978}, but the source of these positrons is still not well understood. 
The 511~keV emission, which is the brightest persistent $\gamma$-ray line in the Galaxy, shows a strong concentration in the GC region with a low-surface brightness contribution from the Galactic disk: a distribution that is unlike anything seen in other wavelengths.
The proposed birth sites of these Galactic positrons include the $\beta^+$ decay of stellar nucleosynthesis products \citep[{e.g. $^{26}$Al, $^{44}$Ti and $^{56}$Ni}][]{prantzos2011, crocker2017, milne1999}, pair production in microquasars, low mass X-ray binaries and millisecond pulsars~\citep{weidenspointner2008,prantzos2006,venter2015, Bartels2018_binaries511}, 
and even dark matter~\citep{boehm2004b}.
However, it is difficult to account for the observational constraints without tuning individual source parameters to extreme values.

Early spectral measurements of the emission found a slightly broadened line at 511~keV, and a low energy continuum from the three-photon decay of ortho-positronium (o-Ps), a short-lived intermediate bound state between a positron and electron~\citep{mohorovici1934,deutsch1951}. 
Significant progress in our understanding of Galactic positrons has been made through spectral studies with the spectrometer SPI~\citep{vedrenne2003} aboard ESA's INTEGRAL satellite~\citep{winkler2003}. 
Using one year of public SPI data, \citet{jean2006} and \citet{churazov2005} performed independent spectral studies of the GC region and found that the annihilation of positrons occurs predominately in warm neutral and partly ionized gas phases in the Galaxy. 
The 511~keV line shape and the o-Ps continuum flux imply the annihilating positrons have kinetic energies at the ``eV'' scale. 
At higher energies, positrons can annihilate in flight to produce a continuous high-energy spectrum; \citet{beacom2006} and \citet{sizun2006} estimated that Galactic positrons cannot have initial energies larger than a few MeV from upper limits of the emission above 511~keV as measured by COMPTEL. The positrons must lose energy and slow down after production, and it is therefore expected that positrons propagate some distance from their production sites to where they annihilate. 
More recently, \citet{siegert2016a} have analyzed more than ten years of SPI data 
and have confirmed earlier results in the spectral domain; however, the authors questioned whether the annihilation conditions, and thus the spectral signatures, are identical throughout the Galaxy.

The imaging analysis, on the other hand, has been less conclusive. 
Initial images of the positron annihilation signature were obtained by the Oriented Scintillation Spectrometer Experiment (OSSE) aboard the Compton Gamma-Ray Observatory (CGRO), launched in 1991. 
By combining OSSE data with scanning observations from the Transient Gamma-Ray Spectrometer (TGRS) and the Solar Maximum Mission (SMM), \citet{purcell1997} produced maps of the emission which showed three distinct features: 1) a central bulge, 2) emission along the Galactic disk, and 3) a positive latitude enhancement. 
This confirmed earlier measurements which detected an enhancement in the GC region and was the first observational evidence of 511~keV emission consistent with the plane of the Galaxy. 
The positive enhancement, which became known as the ``Annihilation Fountain''~\citep{dermer1997}, was not seen in OSSE images of the o-Ps continuum~\citep{milne2001} and is now believed to be an imaging artifact.

Recent attempts to constrain the spatial distribution have been made with SPI observations.
As SPI is a coded-mask imaging telescope, analysis of diffuse emission has relied on a model fitting approach
that makes SPI insensitive to weak gradients and large scale structures much larger than the 16$^{\circ}$ field-of-view (FOV). 
The proposed spatial models remain entirely empirical. 
The first SPI all-sky map of the 511~keV emission \citep{knodlseder2005}, which used a Richardson Lucy deconvolution technique~\citep{richardson1972}, showed a strong emission feature towards the Galactic bulge.
\citet{bouchet2010} analyzed six years of SPI data and found a possible shift of the 511~keV central bulge emission towards negative longitudes, no significant detection of a point source contribution, and they reported on a possible halo emission geometry.
\citet{skinner2014} and \cite{siegert2016a} have performed more recent studies of the spatial distribution and both describe the 511~keV emission in the Milky Way with four empirical 2D Gaussian functions: a narrow and broad bulge with 5.9$^{\circ}$ and 20.5$^{\circ}$ FWHM, respectively, a point source component consistent with Sgr A*, and a disk component. 
The reported extent of the disk emission is drastically different in these two recent studies resulting in a poorly constrained positron annihilation rate in the Galaxy.

After almost five decades of scientific investigation, there are still major questions about Galactic positrons. 
The 511~keV emission from the Galactic disk can potentially be explained by nucleosynthesis products; however, there is no conclusion as to the source of positrons in the Galactic bulge region. 
The spatial morphology of the emission has not been well constrained and it is not clear if the 511~keV emission should trace the distribution of positron sources or if there really is significant positron propagation~\citep{higdon2009, jean2009, alexis2014}.
In addition, there is the question of whether the emission is truly diffuse, as could be expected from gas in the Galaxy, or if there is a large population of unresolved point sources which makes the emission appear smooth.
Both theoretical advancements in the understanding of positron interactions, the constituents of the interstellar medium, and Galactic magnetic fields, and a more accurate image of the 511~keV emission are needed to further advance in this topic.
A direct imaging, wide-FOV telescope would be able to determine the true spatial morphology, conclusively determine the extent of the disk emission, and measure the true annihilation rate from different regions of the Galaxy. 

The Compton Spectrometer and Imager (COSI) is a telescope that has been developed with the goal of furthering our understanding of Galactic positrons.
With its wide FOV and Compton imaging capabilities, COSI, described in Sec.~\ref{sec:instrument}, can shed light on these open questions, especially in regards to morphology.
In this paper, we report on the detection of the 511~keV GC emission with COSI during its 2016 balloon flight from Wanaka, New Zealand, overviewed in Sec.~\ref{sec:observations}.
This is the first detection and characterization of Galactic positron annihilation with a Compton telescope. 
As other telescope technologies rely on distinct background estimation techniques, and thus suffer from different systematics, the COSI measurements provide a unique diagnostic compared to the coded-mask imager SPI and the collimated OSSE. 
To extract the spectral and spatial signature of Galactic positron annihilation, we have developed a technique to estimate the environmental and instrumental background with the COMPTEL Data Space, described in Sec.~\ref{sec:analysis}.
We present a characterization of the the 511~keV line shape and o-Ps continuum fraction, as well as the flux of the 511~keV line, as detailed in Sec.~\ref{sec:spectralresults} and \ref{sec:fluxresults}.
In Sec.~\ref{sec:spatialresults} we present a measurement of the spatial distribution of the Galactic bulge emission with COSI. 
A discussion of the results presented here is found in Sec.~\ref{sec:discussion} and Sec.~\ref{sec:conclusions} is the conclusion of this study.

\section{The Compton Spectrometer and Imager} \label{sec:instrument}

COSI is a soft $\gamma$-ray (0.2--5~MeV) telescope designed to fly on NASA's new 18~million~cubic-foot Super Pressure Balloon (SPB) platform. 
COSI had a successful 46-day balloon flight in May-July 2016~\citep{kierans2016}, and here we present the analysis of the Galactic positron annihilation signature as detected during that flight.

The heart of COSI is composed of twelve cross-strip, high-purity germanium detectors \citep[GeDs;][]{amman2007}. 
Each detector is 8~cm~$\times$~8~cm~$\times$~1.5~cm, where 37 strip electrodes deposited orthogonally on each side give the detectors internal position sensitivity with a 3D voxel size of 1.5~mm${^3}$~\citep{lowell2016}. 
The use of GeDs gives COSI a high spectral resolution of 0.6\% FWHM at 662~keV. 

The twelve detectors are stacked in a 2$\times$2$\times$3 configuration in an aluminum cryostat and have a total active volume of 972~cm$^3$. Cesium iodide scintillators surround the cryostat on the four sides and bottom to act as an anti-coincidence shield to veto background radiation, predominately from Earth's albedo, and constrain the field of view to $\sim\pi$~sr.
The COSI cryostat is fixed on top of a non-pointing, zenith-oriented gondola frame and operates as a free-floating survey instrument.

The COSI instrument has notable heritage. 
Prior to the 2016 flight, the same instrument was flown as a SPB Mission of Opportunity in 2014 from McMurdo Station, Antarctica; unfortunately, the balloon developed a leak and the flight only lasted 43 hours. 
The precursor instrument to COSI, the Nuclear Compton Telescope, saw three previous launches. 
For a description of the COSI instrument and its history, see \citet{bandstra2011} and \citet{kierans2016}, and reference \citet{lowell2017b,lowell2017a} for GRB science results from the 2016 COSI flight.

\subsection{Compton Telescope Basics}

Taking advantage of the dominant interaction mechanism at MeV energies, Compton telescopes use the interaction position and energy deposits in a sequence of Compton scatters to determine the initial photon's energy and source sky position~\citep{vonballmoos1989, boggs2000}. 
In a compact Compton telescope, like COSI, the distance between interactions is too small for time-of-flight methods, and thus the correct temporal order of scatters is determined through Compton Kinematic Reconstruction~\citep{boggs2000}, which uses redundant information in the geometry and kinematics of the scatters to find the mostly likely path of the photon. 

Each event in a Compton telescope is recorded as a measurement of the position and energy of interactions in the detector volume. 
After Compton event reconstruction, the main parameters 
for an event are reduced to the total energy deposited and the geometric angles of the scattered photon in the initial interaction (further discussed in Sec.~\ref{sec:analysis}). These descriptors are used to constrain the initial photon direction to a projected circle on the sky, often called the \textit{event circle}. 
When multiple photons from the same source interact in the detector, the resulting event circles will overlap at the source sky position and iterative deconvolution techniques can be used to create an image.

The angular resolution of a Compton telescope is described by the angular resolution measure (ARM).
The ARM is the smallest angular distance between a known source location and the event circle for each photon. 
The distribution of all ARM values from a sample of Compton events represents the effective width of the point spread function of a Compton telescope. 
Consequently, the FWHM of the ARM distribution defines the achievable angular resolution after event reconstruction.

\subsubsection{Compton Telescope Event Selections}
\label{sec:eventselections}
The source detection significance of a Compton telescope can be improved by rejecting lesser quality events through proper selections. The event selections aim to optimize the signal-to-noise ratio; however, stricter selections will limit the effective area. The selections used in this analysis, and their general effect, are listed below and summarized in Tab.~\ref{tb:511spectrumeventselections}.
\begin{itemize}
 \setlength\itemsep{0em}
    \item The initial Compton scatter angle of the first interaction $\phi$ has a large effect on the quality of the event. Backscatters, with $\phi > 90^{\circ}$, are generally harder to reconstruct, so limiting $\phi$ to less than 90$^{\circ}$ can reject these lower-quality events.
    \item The minimum distance between the first two interactions is the lever arm of $\phi$; a large distance allows for the direction of the $\gamma$-ray to be more accurately determined. The distance between any interaction can also be chosen, where a larger distance will give a more precise reconstruction.
    \item The length of the Compton sequence is the number of allowable interactions. Events with three or more interactions contain redundant information and are easier to reconstruct.
    \item The Earth Horizon Cut (EHC) rejects any event whose Compton event circle overlaps with the horizon. This is the most rigorous method to reduce the background from albedo radiation during flight.
    \item An origin cut can be made on a location in image space with a given radius. Only Compton events which overlap with this origin selection will be kept.
\end{itemize}

\begin{table}[h]
\centering
\begin{tabular}{lc}
\hline 
Parameter										& Allowed Range 			\\
\hline
Altitude										& $\geq$ 27000~m				\\
Origin selection									& 16$^{\circ}$  (if applicable)	\\
Photon energy									& 506 -- 516 keV (if applicable)	\\
Number of interactions							& 2 -- 7   					\\
Compton scatter angle							& 15 -- 55$^{\circ}$			\\
\multirow{2}{3cm}{Distance between first 2 interactions}  	&  $>$ 0.5~cm 				\\ 
											&						 \\
\multirow{2}{3cm}{Distance between any interaction}		& $>$ 0.3~cm 				\\ 
											& 						\\
Earth horizon cut		& \multirow{2}{4.5cm}{Reject if any part of event circle is below horizon} \\
& \\
\hline
\end{tabular}
\caption{Compton telescope event selections for the presented analysis. The origin selection is used only for the spectral subtraction (Sec.~\ref{sec:spectralroutine}), and the photon energy cut is only used for the CDS-ARM subtraction (Sec.~\ref{sec:spatialroutine}).}
\label{tb:511spectrumeventselections}
\end{table}



\subsection{Analysis Package}

The COSI collaboration employs the Medium Energy Gamma-ray Library  \citep[MEGAlib;][]{zoglauer2006} for its primary data analysis pipeline. 
MEGAlib is a set of software tools which specializes in Compton telescope data analysis: applying instrument calibrations, performing Compton reconstruction, and implementing image reconstruction. 
MEGAlib also has tools for accurate instrument simulations based on GEANT4~\citep{geant4}, in which a detailed description of the measured detector performance can be applied to Monte Carlo simulations. 
For a description of the COSI analysis pipeline and a thorough comparison between simulations and laboratory measurements taken prior to the 2016 campaign, see \cite{sleator2019}.

\section{Observations} 
\label{sec:observations}
COSI was launched from Wanaka, New Zealand (45$^{\circ}$~S, 169$^{\circ}$~E), on May 17, 2016 (23:35 05/16/16 UTC) on-board NASA's  SPB platform. 
The flight was terminated after 46 days and the instrument landed 200~km north-west of Arequipa, Peru, on July 2, 2016 (19:54 07/02/16 UTC; 16$^{\circ}$ S, 72$^{\circ}$ W). 
The trajectory of the instrument covered a range of latitudes from 60$^{\circ}$~S to 6$^{\circ}$~S and included a total circumnavigation of the Earth. 
The nominal float altitude was 33~km; however, large altitude drops occurred at night during the latter half of the flight due to anomalies in the balloon. 
Three of the twelve GeDs had high-voltage related issues during the flight: two were non-operational after 48 hours and one failed 20 days into the flight. 
The loss of these detectors decreased the effective area by close to 50\% for these observations; the COSI team has since determined and corrected the high-voltage issues for future flights.
A more detailed description of the COSI 2016 flight can be found in \citet{kierans2016}.

\begin{figure}
    \centering
    \includegraphics[width=8.5cm]{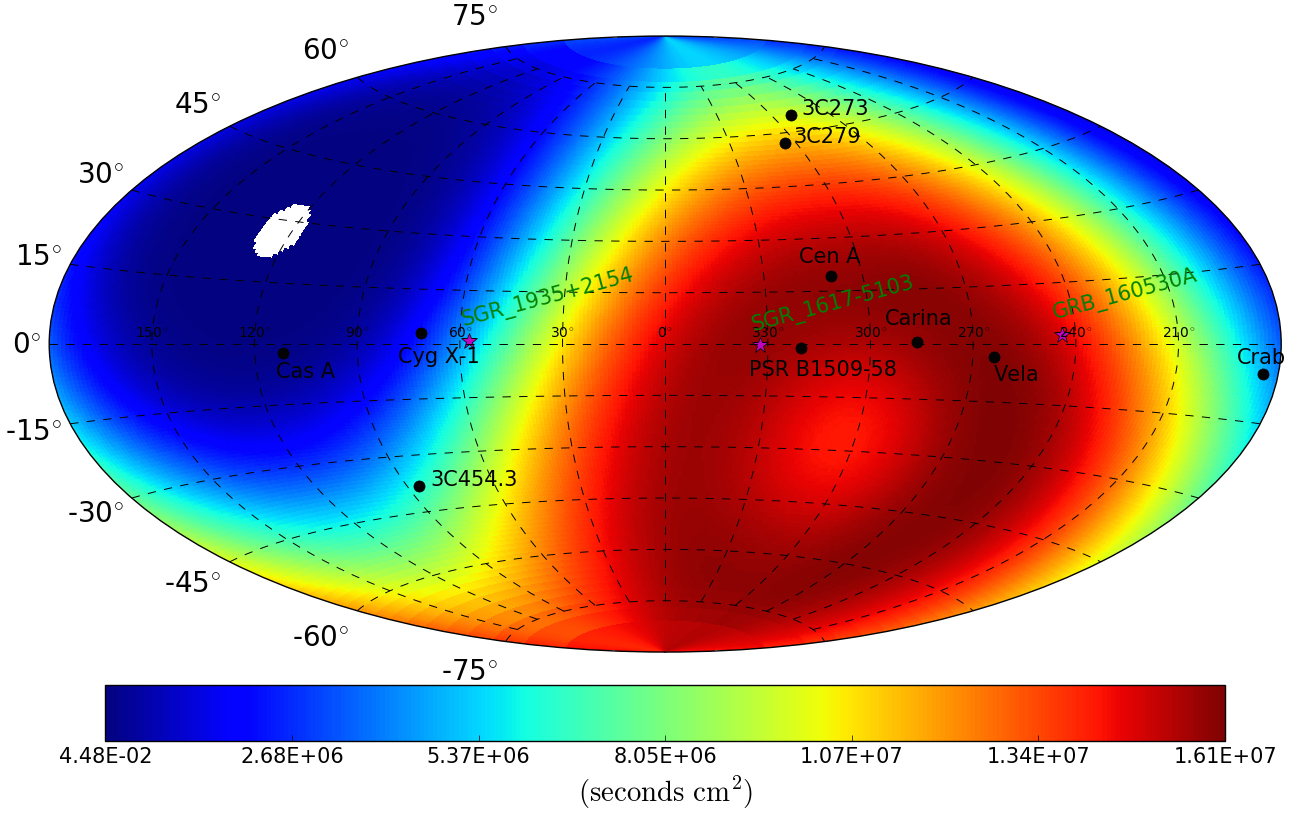}
    \caption{Exposure map from the 2016 COSI flight assuming an effective area of 20 cm$^{2}$. COSI had excellent exposure of the GC, the eastern side of the Galactic plane, as well as the Galactic south pole.}
    \label{fig:exposure}
\end{figure}

\begin{figure}
    \centering
    \includegraphics[width=9.2cm]{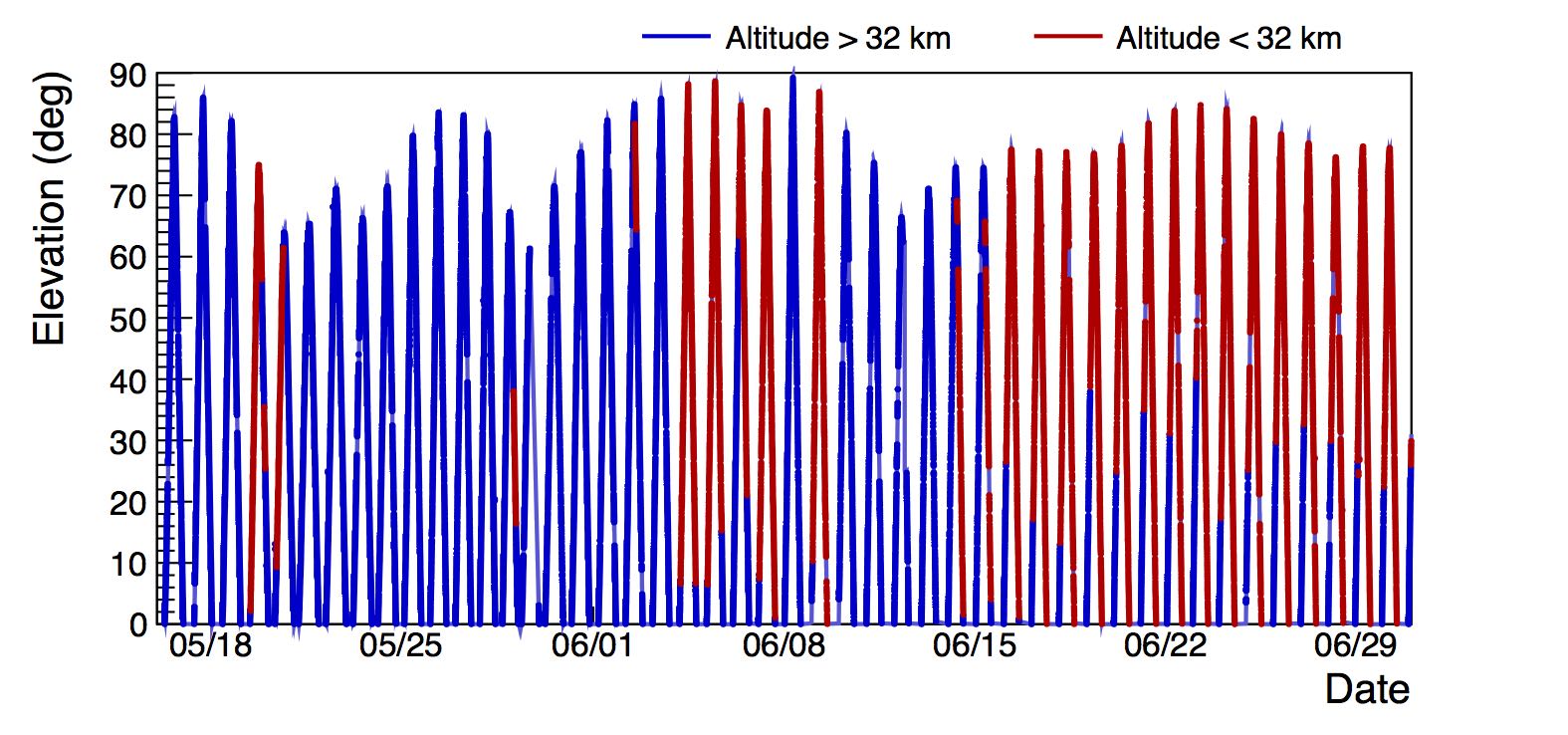}
    \caption{Elevation of the GC in COSI's FOV during the 2016 flight. An elevation of 90$^{\circ}$ degrees occurs when the GC is directly overhead and consistent with COSI's zenith. The times in which the altitude dropped below 32~km are indicated in red.}
    \label{fig:elevation}
\end{figure}

Southern latitudes provide excellent exposure of the GC region, which is necessary for Galactic 511~keV studies. 
From the 2016 flight, COSI had a total of 1.6~Ms of exposure of the GC region, considering times when the GC was within 40$^{\circ}$ of COSI's zenith. 
Figure~\ref{fig:exposure} shows the full flight exposure map with prominent $\gamma$-ray sources labeled. 
Figure~\ref{fig:elevation} shows the elevation of the GC for the duration of the flight, where 90$^{\circ}$ corresponds to COSI's zenith and 0$^{\circ}$ represents the horizon. 
This figure also indicates the times when the altitude of the payload descended below 32~km, depicted in red. 
Unfortunately, the GC was in COSI's FOV only during the night hours, when the altitude eventually began to drop due to the lower atmospheric temperatures. 
At lower altitudes, there is more attenuation of $\gamma$-rays in the atmosphere which directly impacts the observation. 
At the expected float altitude of 33~km, the nominal transmission probability at 500~keV is 49\%, but, for example, at 27~km the transmission probability is only 18\% at zenith.
At the nominal float altitude of 33~km, the average transmission probability for a 500~keV source at the GC is 46\%, where we have included the flight aspect information when the elevation of the GC is above 40$^{\circ}$. This is reduced to an average of 30\% transmission when accounting for the loss of altitude.
The total GC exposure time when the altitude was above 33~km is then reduced to 610~ks.

\begin{figure}
    \centering
    \includegraphics[width = 8.9cm]{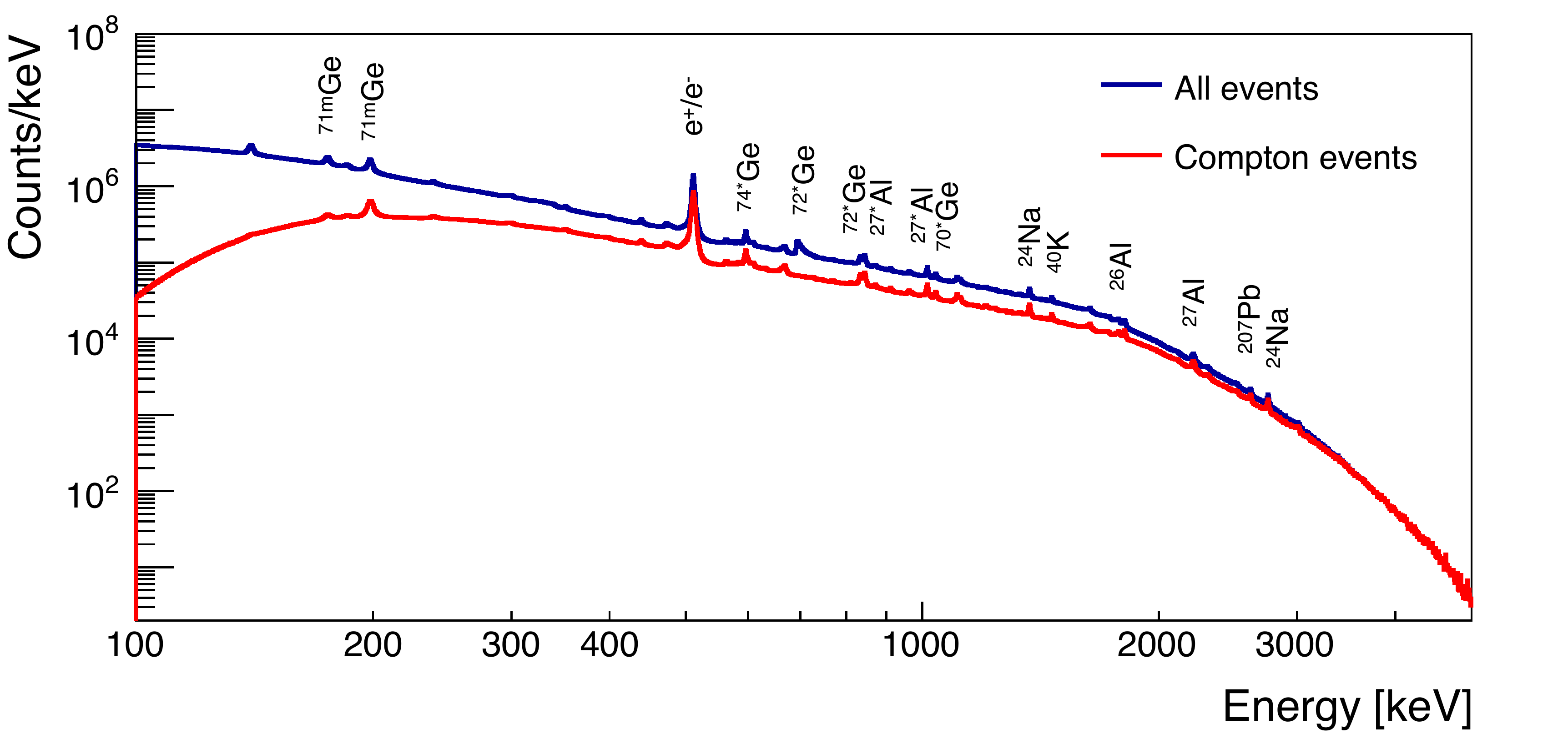}
    \caption{Total integrated energy spectrum, including single-site and multi-site events, from the duration of the COSI 2016 flight. The intense 511~keV atmospheric background line and known activation lines have been identified.}
    \label{fig:backgroundspectrum}
\end{figure}

One of the difficulties of MeV $\gamma$-ray astrophysics is the dominant background radiation. 
Figure~\ref{fig:backgroundspectrum} shows the total accumulated spectrum from the flight. 
The majority of photons in this spectrum are from atmospheric emissions, i.e., $\gamma$-rays from cosmic-ray interactions in the atmosphere. 
Furthermore, when the instrument is bombarded with protons, neutrons and other cosmic-ray particles in the upper atmosphere, nuclear reactions will be induced within the instrument material. 
Radioactive isotopes that have a half-life longer than the timing resolution of the detector, but less than the flight duration, will eventually decay and could appear as a Compton event. 
Some of the activation lines, which are mostly from germanium, are labeled in Fig.~\ref{fig:backgroundspectrum}.
The 511~keV line has background contributions from both activation and atmospheric emissions and the intensity of these background components  predominately depends on the atmospheric depth, geomagnetic cutoff rigidity, and zenith angle~\citep{ling1977}.

To extract the Galactic positron annihilation spectrum, a precise description of the observed background radiation is required.  
A physical, standalone, background model is not able to capture all the observed dynamics during the flight.
However, calculating the variations of the background as a function of Earth longitude and latitude, resulting in rigidity values, as well as taking into account the flight altitude at every instance in time results in large systematics, especially for the 511~keV background line. 
Therefore, a sophisticated technique to separate out the Galactic emission from the from the background has been developed. 
In particular, the recorded events are analyzed according to their expected appearance in the fundamental data space of Compton telescopes. 
This is detailed in Sec.~\ref{sec:analysis}.

For the analysis presented here, we use all 46 days of flight, excluding times when the background rates were high due to electron precipitation events~\citep{millan2007, kierans2016} and times when the altitude dropped below 27 km. 

\section{Analysis Method}
\label{sec:analysis}
The COMPTEL collaboration pioneered the analysis tools for Compton telescopes. 
In particular, they performed the majority of their analyses in a three-dimensional data space, referred to here as the COMPTEL Data Space (CDS). 
We utilize the CDS background handling approach as introduced for the $^{26}$Al $\gamma$-ray line measurement with COMPTEL~\citep{knodlseder1996}. 
To perform an accurate estimate of the background for the 511~keV line in COSI data, we further developed this technique.
This is the first instance that this method has been applied for a compact Compton telescope, like COSI, and for Galactic positron annihilation analysis.
A similar approach for analysis of continuum point sources detected by COSI is presented in \citet{sleatorthesis2019}.

\begin{figure}
    \centering
    \subfloat[Compton event with source at ($\chi_0,\phi_0$).]{\includegraphics[width = 7cm]{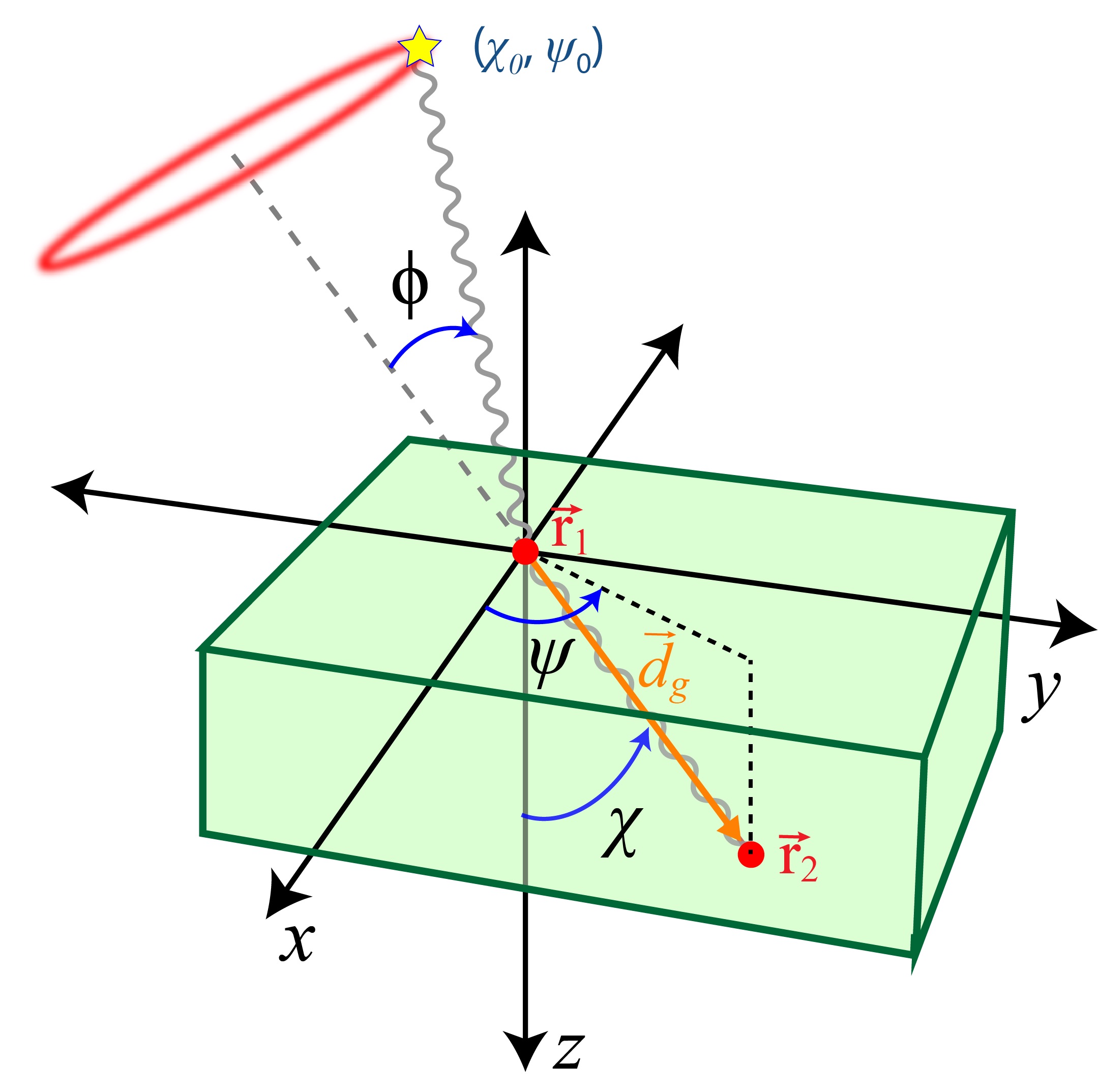}}
    \hfill
    \subfloat[Accumulation of events in the CDS with source at ($\chi_0$, $\phi_0$).]{\includegraphics[width = 4.2cm]{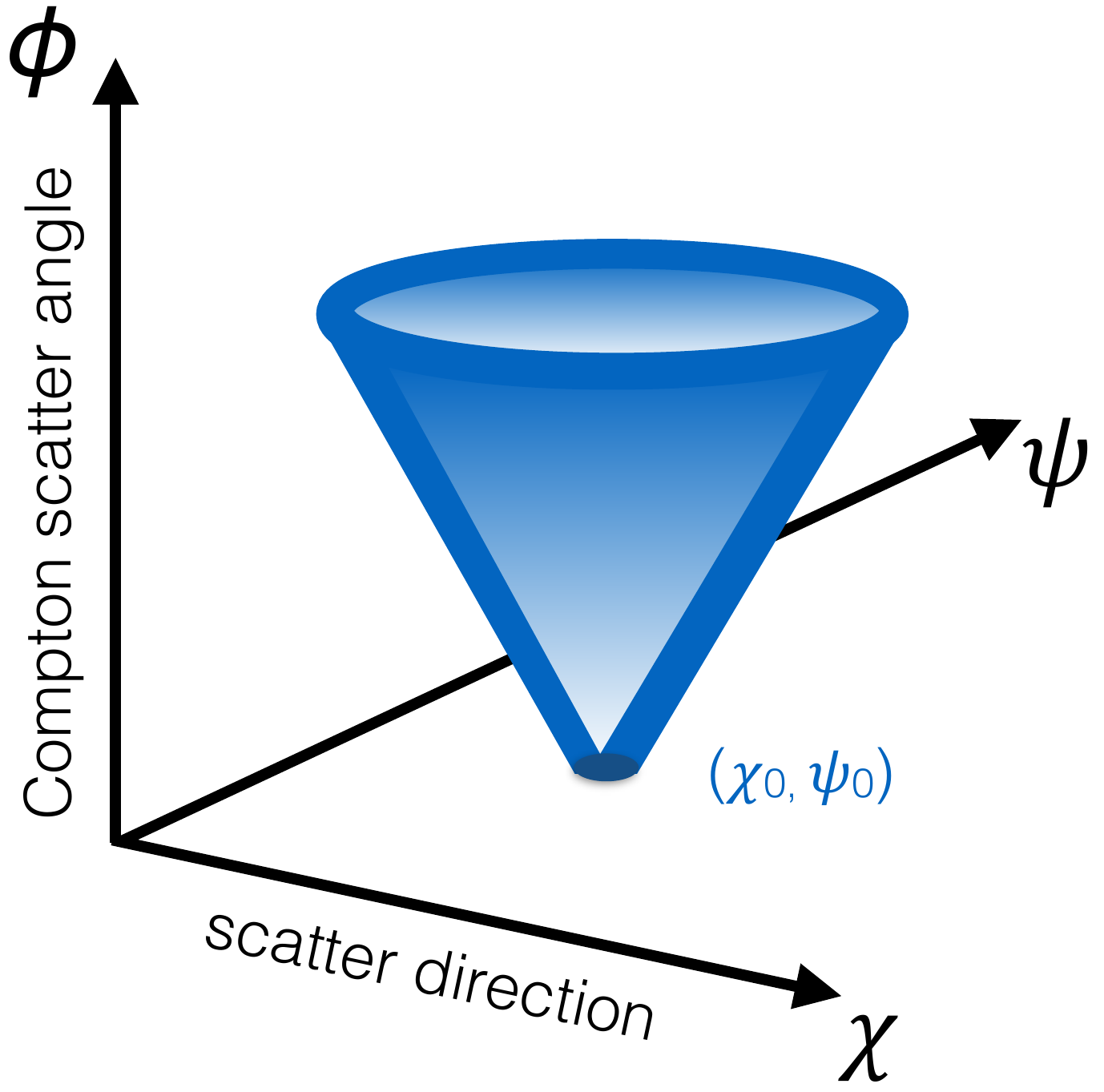}}
    \hfill
    \subfloat[CDS for an on-axis source.]{\includegraphics[width = 4.2cm]{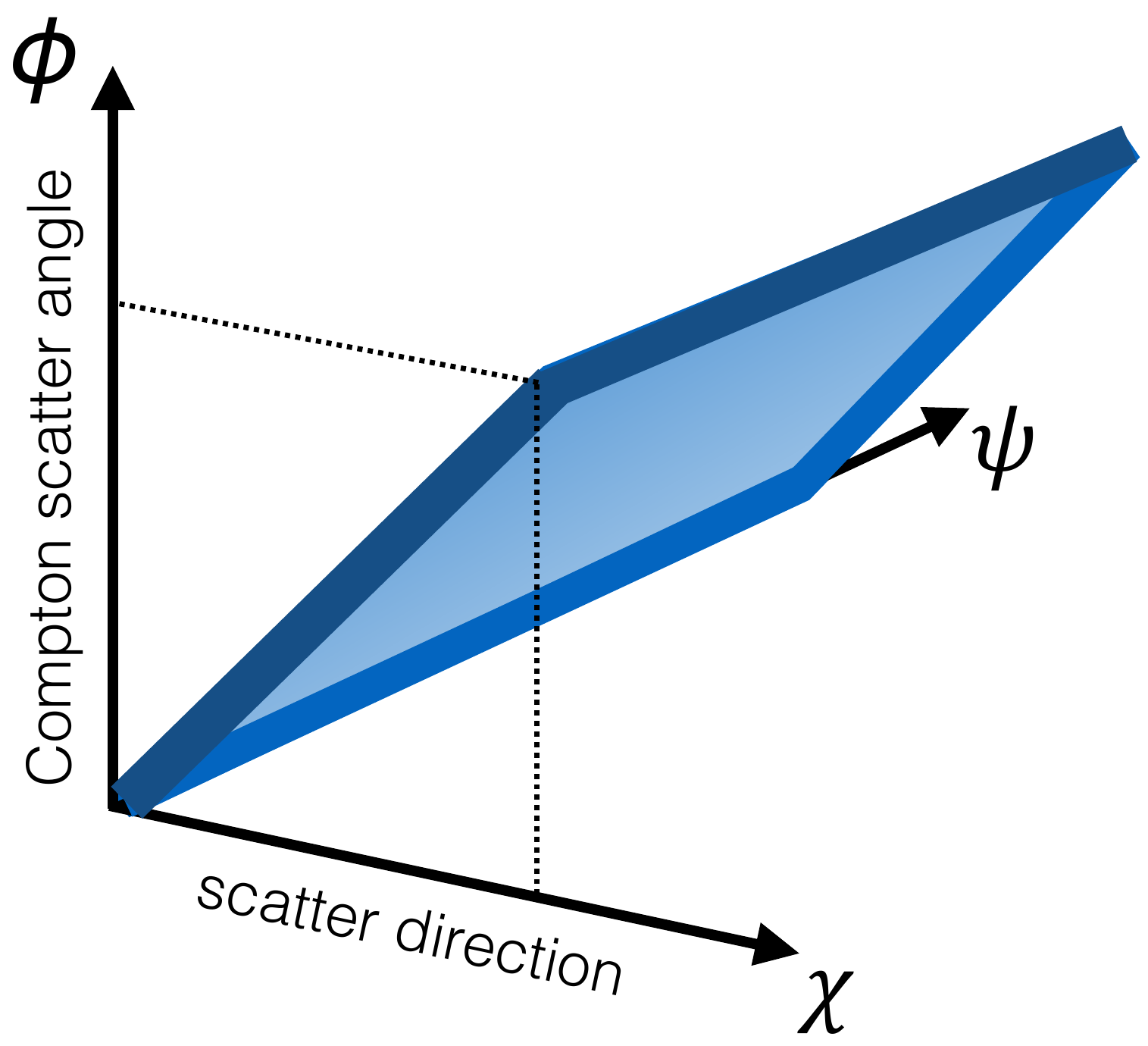}}
    \caption{(a) Schematic diagram of the first two interactions of a Compton event showing the CDS angles. The source is at $(\chi_0, \psi_0)$, and in the far field the radius of the event circle in image-space (red) is equal to the Compton scatter angle $\phi$, which defines one of the axes of the CDS. The polar and azimuthal angles, $\chi$ and $\psi$, of the Compton-scattered $\gamma$-ray direction $\vec{d_g}$ define the two other axes of the CDS. (b) Once many photons from the same source are accumulated, with each event represented as a point at $(\chi, \psi, \phi)$, the source is mapped as a cone in the CDS with its apex at the source position $(\chi_0, \psi_0)$. (c) When a source is on-axis, or at the GC in Galactic coordinates, the CDS cone is transformed into a plane along $\chi = \phi$.}
    \label{fig:COMPTELDSAxes}
\end{figure}

The three orthogonal axes of the CDS are defined by the polar and azimuthal angles, $\chi$ and $\psi$, respectively, of the first Compton scatter direction $\vec{d_g}$ in detector coordinates, and the Compton scatter angle $\phi$ of the first interaction. 
In Fig.~\ref{fig:COMPTELDSAxes}~(a), a schematic illustrates the definition of the three CDS angles for a single event. 
The total energy of the $\gamma$-ray can be considered the fourth dimension of this data space.
In contrast to the projected event circle in image-space, each Compton event is a point in the CDS at $(\chi, \psi, \phi)$. 
The accumulation of Compton events from point-source emission in the instrument's FOV, depicted by the yellow star at $(\chi_0, \psi_0)$ in Fig.~\ref{fig:COMPTELDSAxes}~(a), populates the surface of a 3D cone in the CDS; see Fig.~\ref{fig:COMPTELDSAxes}~(b).
The CDS cone has its apex at the source position $(\chi_0, \psi_0)$ in detector polar coordinates because in the limit that $\phi \rightarrow 0$, $\vec{d_g}$ will point towards the source location.
The opening angle of the cone is 90$^{\circ}$ since the Compton scatter angle and polar scatter direction increase at the same rate.
For a point source, the surface of the cone is as thick as the angular resolution of the instrument, while extended sources, such as the Galactic 511~keV emission, produce a thicker cone and an extended apex.

For flight observations, we convert the $\chi$ and $\psi$ dimensions of the CDS from detector coordinates into Galactic coordinates with known aspect information of COSI for every Compton event. 
With the CDS in Galactic coordinates, a source at the GC would put the apex of the CDS cone at the origin of the data space and the cone shape is transformed into a 2D plane 45$^{\circ}$ relative to the $\phi$ and $\chi$ axes, see Fig.~\ref{fig:COMPTELDSAxes}~(c).
The azimuthal scatter direction $\psi$ in detector coordinates is now equal to the Klein-Nishina azimuthal scatter angle~\citep{nishina}, which encodes the polarization of the incoming emission. 
However, the Galactic positron annihilation emission is not expected to be polarized and, therefore, we can integrate over the $\psi$ dimension with no loss of information.
The CDS is then projected into a 2D plane defined by $\chi$ and $\phi$.

\begin{figure}
\centering
\subfloat[On-axis point source simulation.]{\includegraphics[width=8cm]{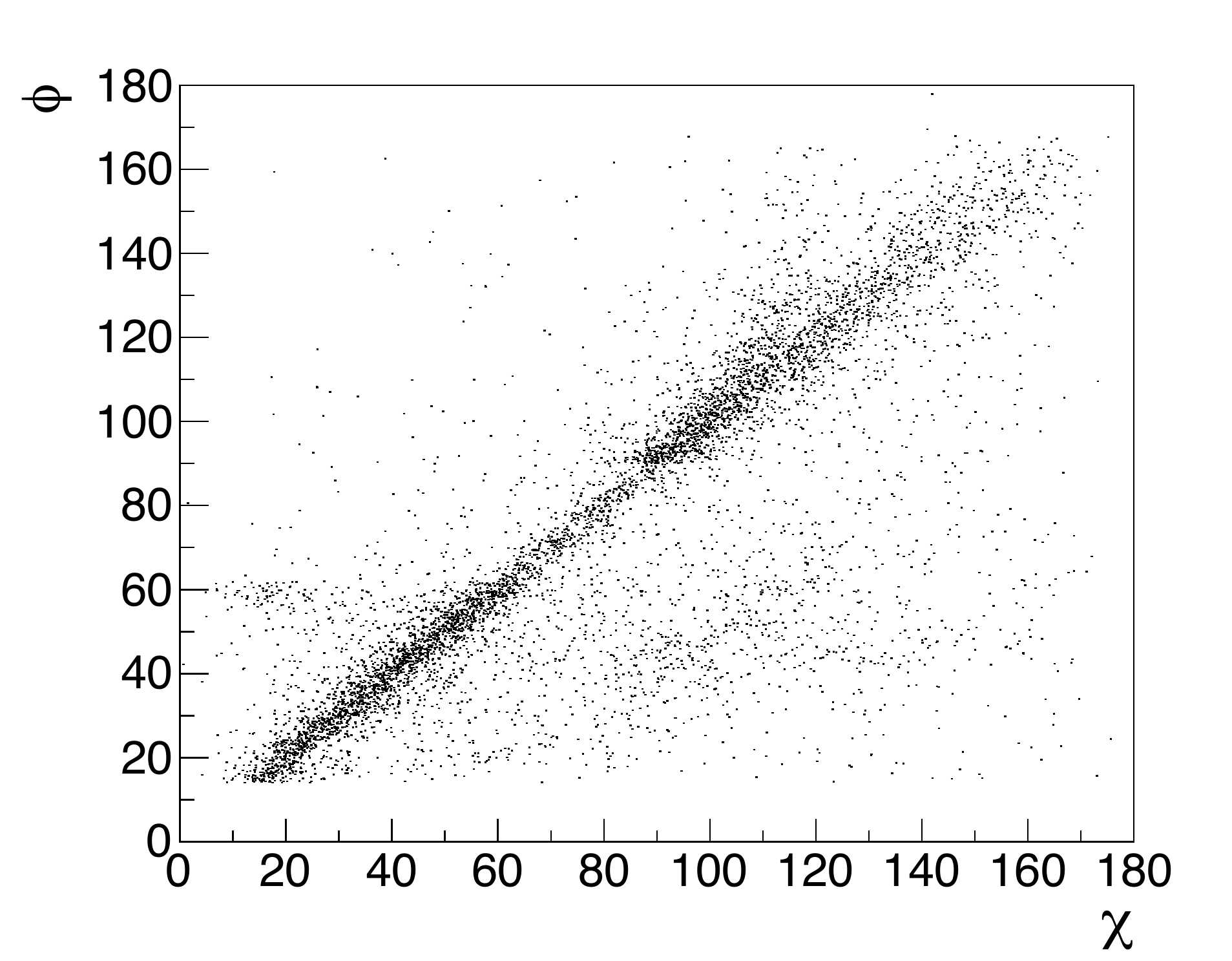}}\\
\subfloat[CDS-ARM of point source simulation.]{\includegraphics[width=8cm]{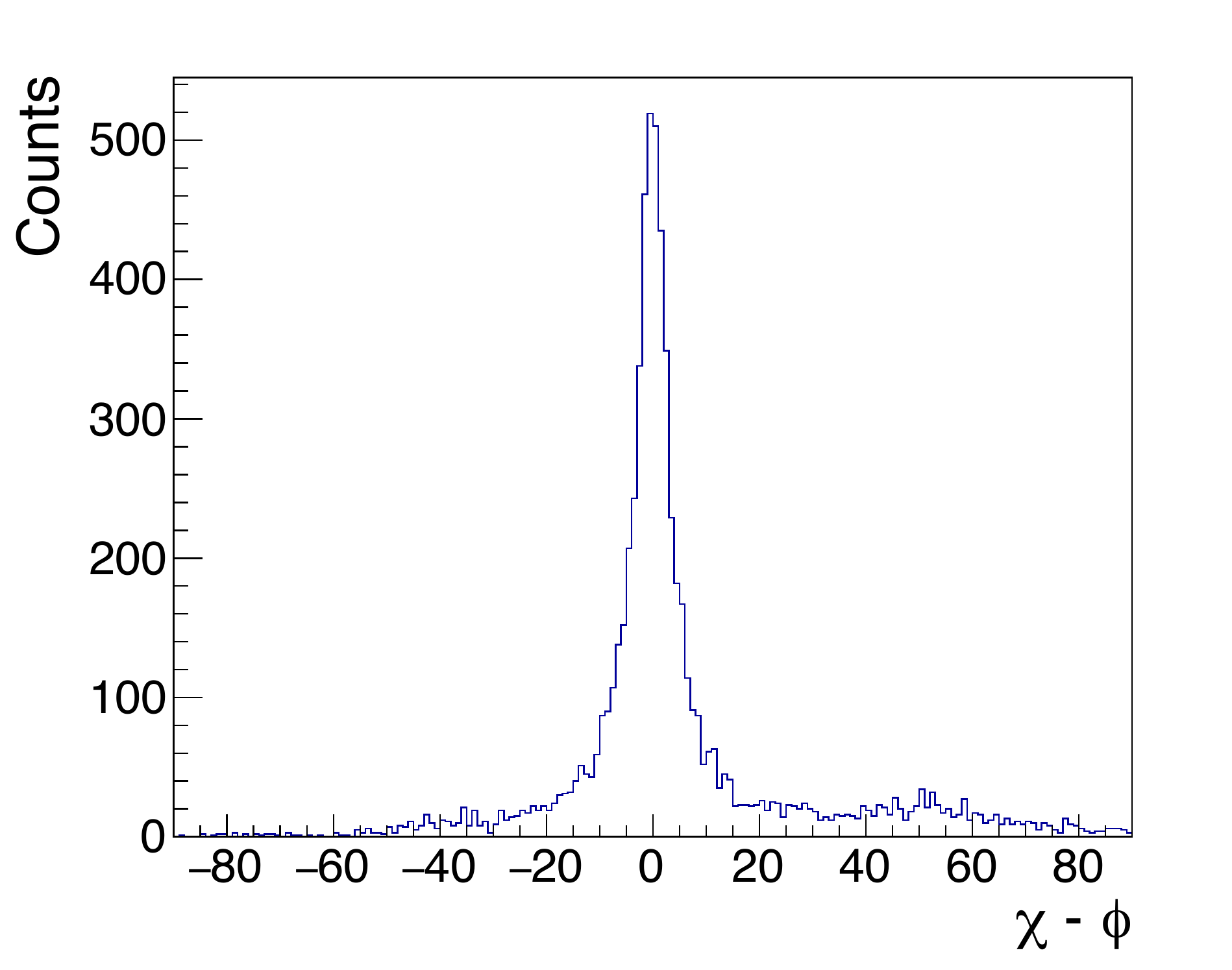}}
\caption{(a) Simulation of a 511~keV point source at (0, 0) mapped in the CDS. Only total $\gamma$-ray energies between 506--516~keV are included so as to select on the fully absorbed and properly reconstructed events to best represent the angular resolution. (b) The deviation of events from the true source location line at $\chi = \phi$, given by the distribution of $(\chi - \phi)$ shown here for the point-source simulation, defines the angular resolution of the telescope. This radial distribution around the source location is} referred to as the CDS-ARM and is equivalent to the ARM distribution defined in image-space.
\label{fig:2DCDSandARM}
\end{figure}

Ideally, a source at the GC in this 2D space is mapped to a line at $\chi = \phi$.
The resulting 2D CDS representation of a simulated 511~keV point source observed at COSI's zenith is shown in Fig.~\ref{fig:2DCDSandARM}~(a); we have made an energy selection of 506--
516~keV to select on the fully absorbed events.
Due to the finite energy and position resolution in our detectors and Doppler broadening~\citep{zoglauer2003}, there is a spread to this distribution. 
The deviation from the ideal $\chi = \phi$ line is equivalent to the ARM distribution, i.e., the effective point spread function. 
We define the distance of each event from the $\chi = \phi$ line as $\chi - \phi$, as opposed to the closest distance to the line, given by $(\chi - \phi)/2$.
We will refer to this angular distance as the CDS-ARM. 
The CDS-ARM histogram of the on-axis point source simulation is shown in Fig.~\ref{fig:2DCDSandARM}~(b). 
The CDS-ARM is the radial distribution of events around the source position, and we will used these terms interchangeably. 
If the source is extended, then the radial distribution will be broadened.
The use of this reduced 2D CDS is generalizable for sources not at the GC by rotating any source location into the origin of the CDS; see \citet{kieransthesis2018} for details.

\begin{figure}
\centering
\includegraphics[width = 7cm]{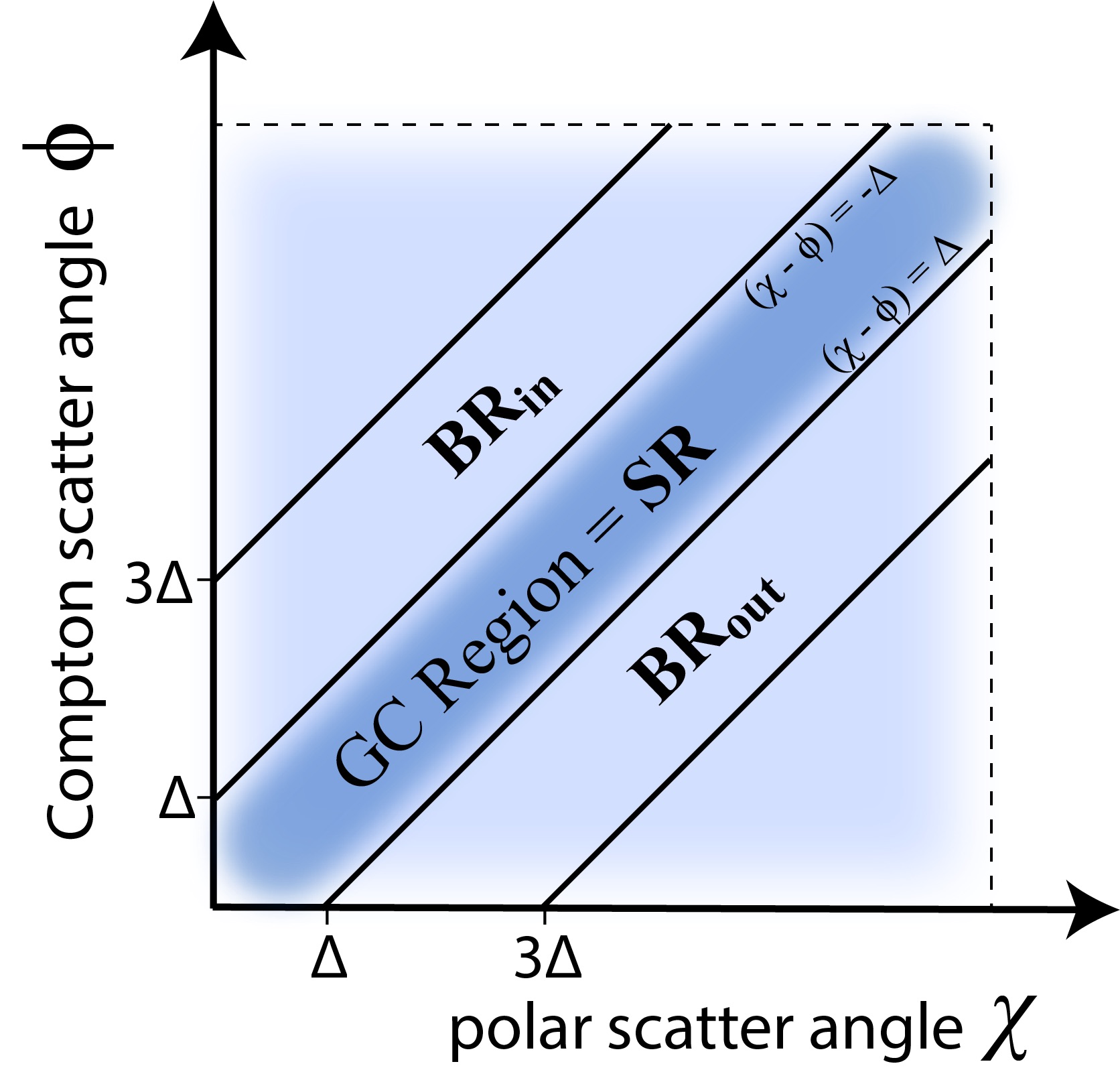}
\caption{Schematic diagram of the 2D CDS showing the source and background regions, where the GC region is shaded darker blue to indicate a larger number of counts from the source. $\mathbf{SR}$ is defined by an origin cut of $\Delta$. In the 3D CDS, the background regions are chosen as concentric cones that surround and sit within the source cone. In 2D, these cones are transformed into lines that lie adjacent to the source line at $\chi = \phi$.}
\label{fig:backgroundregions}
\end{figure}

In general, we can define a region of interest, or source region $\mathbf{SR}$, for these observations as an angular cut around the GC, and therefore a cut around the $\chi = \phi$ line. 
In Fig.~\ref{fig:backgroundregions} we illustrate this 2D CDS with the source region $\mathbf{SR}$ defined as an origin cut (see Sec.~\ref{sec:eventselections}) of $\pm \Delta$ around the GC, and the background regions $\mathbf{BR_{in}}$ and $\mathbf{BR_{out}}$ as adjacent cuts.

The relative population of the background regions depends on $\phi$. 
For this analysis, we find that an origin cut of $\Delta=16^{\circ}$ results in the largest signal-to-noise ratio. 
We also find that the significance of the detection is larger when only $\mathbf{BR_{out}}$ is used.
This is understood to be due to the geometrical differences in the background regions; $\mathbf{BR_{in}}$ does not begin to be populated until $\phi \geq \Delta$, as can be seen in Fig.~\ref{fig:backgroundregions}. 
Since smaller Compton scatter angles allow for a more accurate reconstruction, the $\mathbf{BR_{out}}$ region is more suited to determine the background spectrum in the analysis presented here, where we are selecting only the Compton scatter angles in the range 15--55$^{\circ}$. Therefore, we only use $\mathbf{BR_{out}}$ to estimate the background for this analysis. 

\subsection{Spectral Background Estimation Routine}
\label{sec:spectralroutine}

Although the Compton scatter angle $\phi$ is strongly dependent on energy, in this analysis, we take advantage of the fact that the scatter angle direction $\chi$ is energy-independent. 
For each Compton scatter angle $\phi$ bin, we fill the CDS and find the spectrum from the source in  $\mathbf{SR}$ and the surrounding background region $\mathbf{BR}$.
To account for the fact that the source and background regions of the CDS are not evenly populated, we estimate an ``off-measurement'' by scaling the background region spectrum for an adjacent energy range that contains no source contribution; here we use 520--720~keV for the positron annihilation emission.

The following four-step process defines how the relevant background in COSI data is estimated in the CDS:
\begin{enumerate}

\item Fill the CDS with all events. 
For $\mathbf{SR}$ and $\mathbf{BR}$ separately, bin the Compton scatter angle in the CDS. The measurement statistics allows us to define $\phi$ bins as small as 1$^{\circ}$.
For an origin cut of $\Delta$, the spectrum for the region consistent with $\mathbf{SR}$ in $\phi$ bin $i$ is
\begin{equation}
N^{\text{SR}}(\phi_i, E ) = \sum_{\chi = \phi - \Delta}^{\phi + \Delta} n(\chi, \phi, E),
\end{equation}
where $n(\chi,\phi,E)$ is the number of counts in the $(\chi, \phi, E)$ bin of the CDS. These are the ``on-source'' spectra.
The $\phi$-dependent spectra for the outer background region $\mathbf{BR_{out}}$ in bin $i$ are
\begin{equation}
N^{\text{BR}}(\phi_i, E) = \sum_{\chi = \phi + \Delta}^{\phi + 3\Delta}  n(\chi, \phi, E).
\end{equation}

\item Find the scaling factor for each background spectrum $N^{\text{BR}}(\phi_i,E)$ so that the number of counts in a higher-energy range, 520--720~keV for these studies, equals that in $N^{\text{SR}}(\phi_i,E)$ within the same energy band. The scaling factor, $F_{\phi_i}$, for each Compton scatter angle bin then is:
\begin{equation}
 F_{\phi_i} = \frac{N^{\text{SR}}(\phi_i, E| 520<E<720)}{N^{\text{BR}}(\phi_i, E| 520<E<720)}.
\end{equation}

\item For each $\phi$ bin $i$, scale the background region spectrum $N^{\text{BR}}(\phi_i, E)$ by $F_{\phi_i}$, to obtain a background estimate,
\begin{equation}
\label{eq:CDSphiscale}
B^{\text{SR}}(\phi_i, E) =  F_{\phi_i} N^{\text{BR}}(\phi_i, E),
\end{equation}
i.e., an estimate for an ``off-measurement.''

\item With our ``on-source'' measurements $N^{\text{SR}}(\phi_i, E )$ from Step~1 and our background estimate $B^{\text{SR}}(\phi_i, E)$ from Step~3, we can find the source spectrum by summing the remaining counts in each $\phi$ bin: 
\begin{equation}
S(E) = \sum_i \left [ N^{\text{SR}}(\phi_i, E) - B^{\text{SR}}(\phi_i, E) \right ].
\end{equation}
\end{enumerate}

\begin{figure}
    \centering
    \subfloat[{$\mathbf{SR}$ and $\mathbf{BR}$ spectra for $\phi=\left[ 20^{\circ},21^{\circ} \right]$.}]{\includegraphics[width = 9cm]{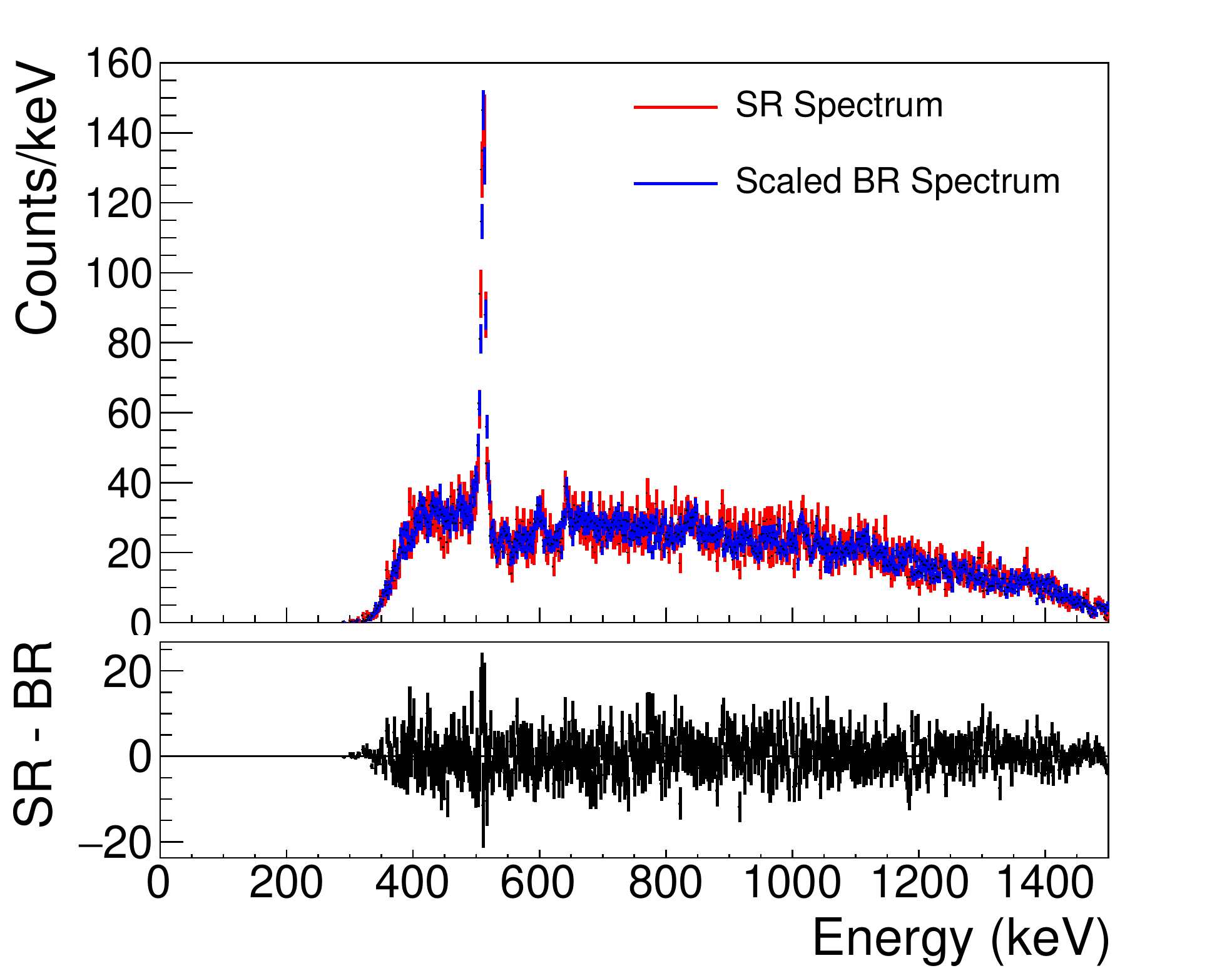}}
    \hfill
    \subfloat[{$\mathbf{SR}$ and $\mathbf{BR}$ spectra for $\phi = [40^{\circ},41^{\circ}]$.}]{\includegraphics[width = 9cm]{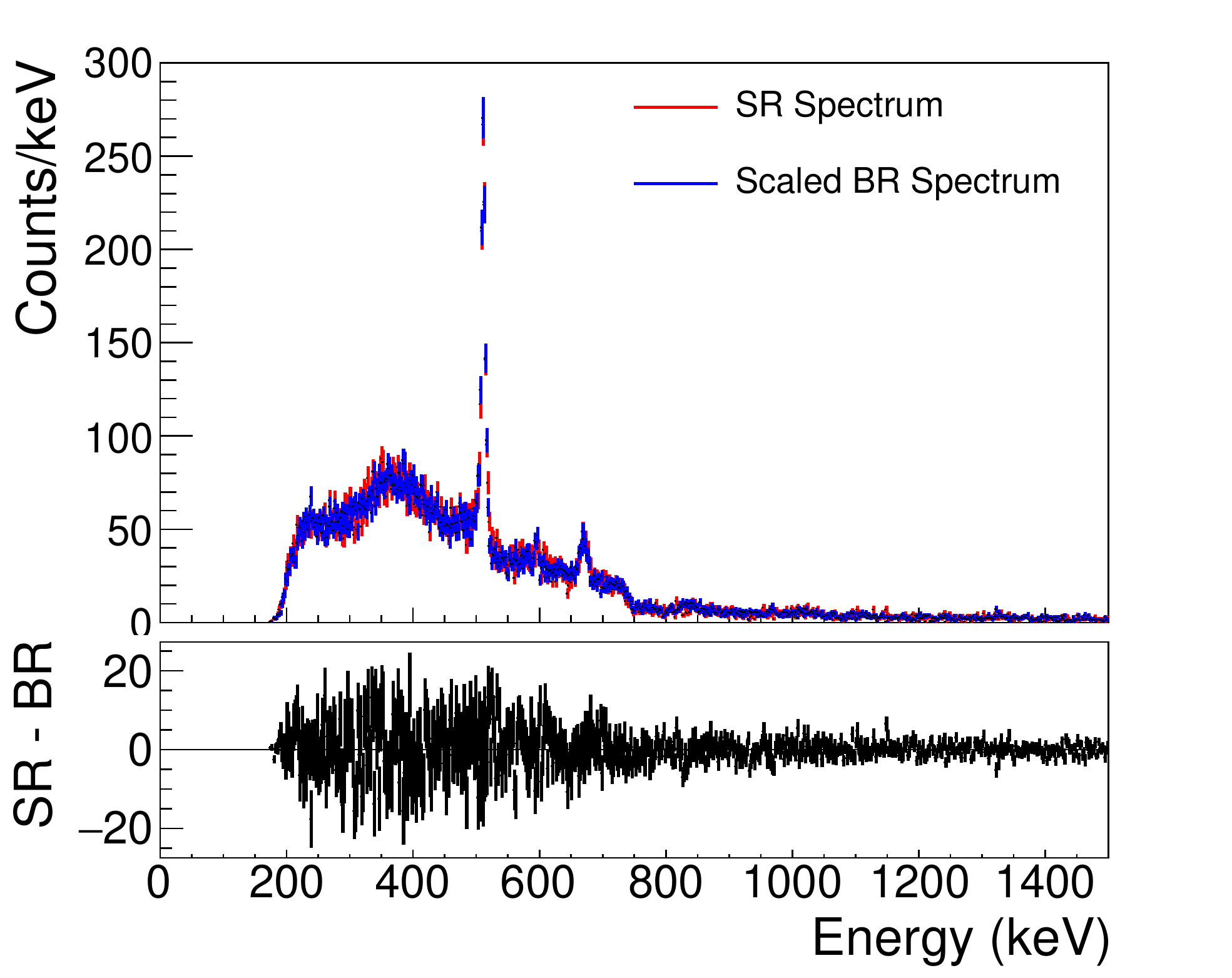}}
    \caption{(a) Spectrum of measured flight data in the source region with Compton scatter angle $\phi = [20^{\circ},21^{\circ}]$. The spectrum from the background region, shown in blue, has been scaled so the number of counts between 520-720~keV is equal to that in the source spectrum. The residuals of the $\mathbf{SR}$ minus $\mathbf{BR}$ is shown in the lower panel. Error bars are taken to be $\sqrt{N}$ of the bin counts. (b) Same as above but for Compton scatter angle $\phi = [40^{\circ},41^{\circ}]$.}
    \label{fig:spectrumslices}
\end{figure}

\begin{figure}
    \centering
    \subfloat[{$\chi$ distribution for $\phi =[20^{\circ}, 21^{\circ}]$.}]{\includegraphics[width = 8cm]{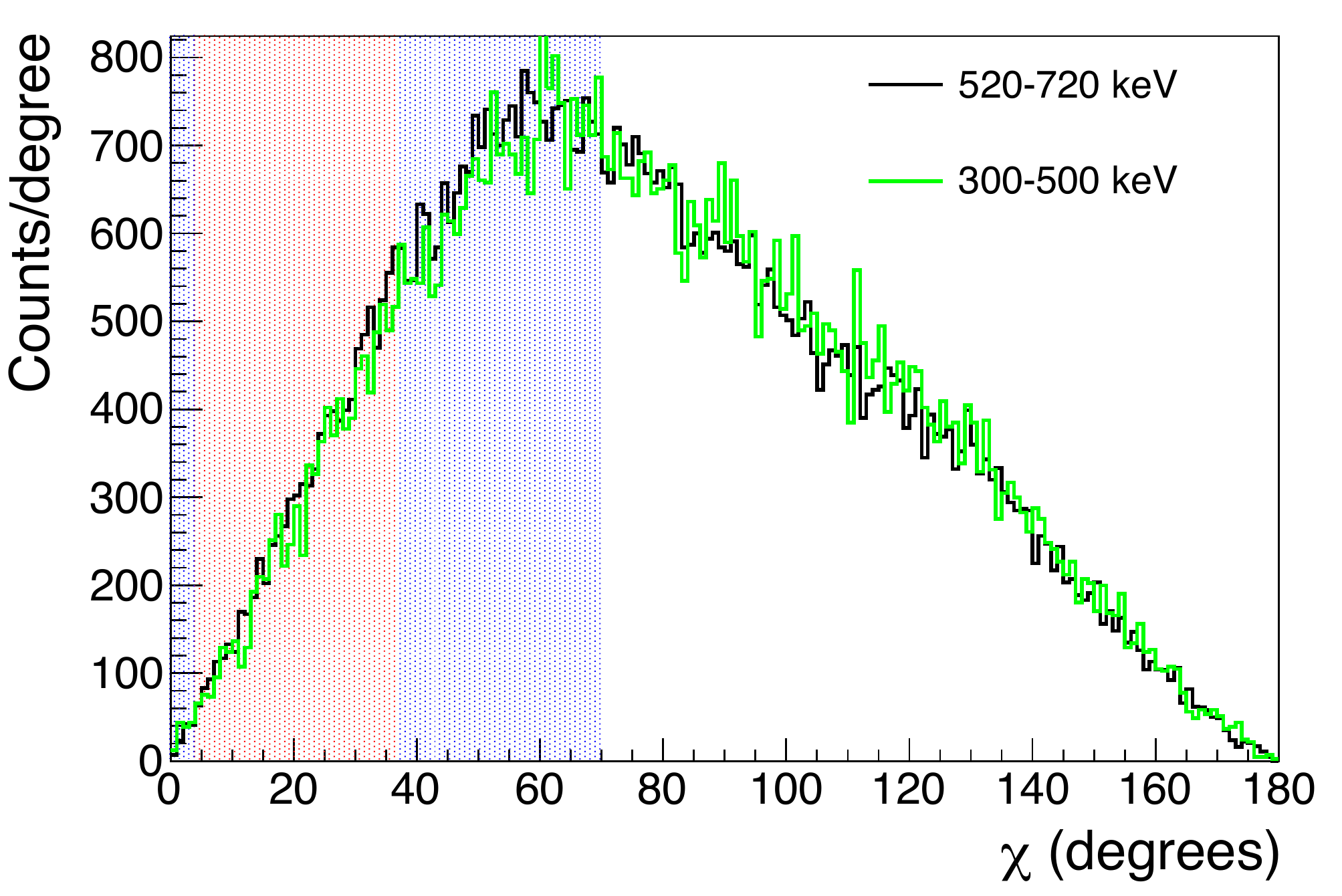}}
    \hfill
    \subfloat[{$\chi$ distribution for $\phi =[40^{\circ},41^{\circ}]$.}]{\includegraphics[width = 8cm]{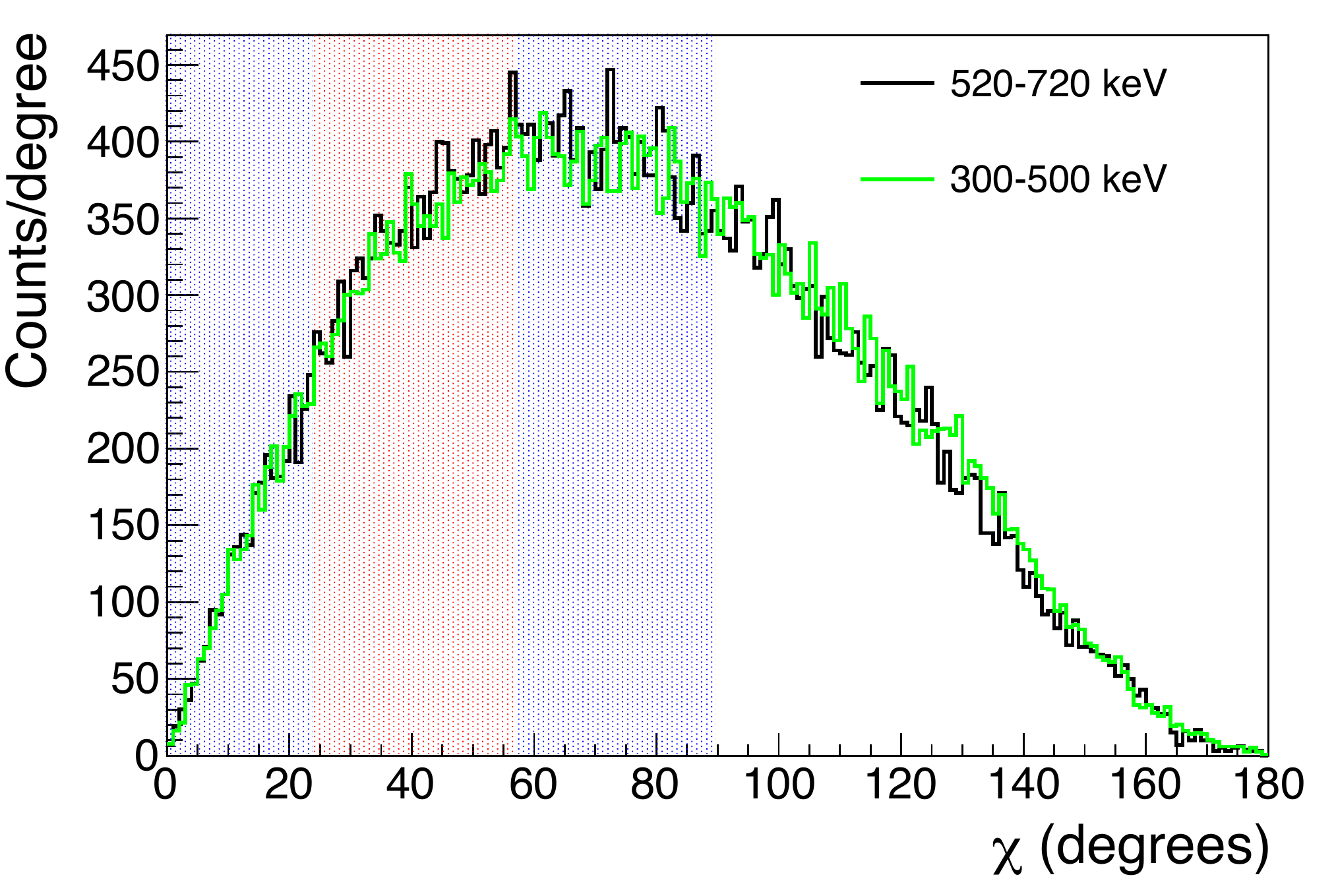}}
    \caption{(a) Using simulated flight data, the $\chi$ distribution for Compton scatter angles $\phi = [20^{\circ},21^{\circ}]$ is compared for two energy ranges: 300--500 keV and 520--720~keV.  
    The red and blue regions signify the location for $\mathbf{SR}$ and $\mathbf{BR}$, respectively, assuming an origin cut of $\Delta = 16^{\circ}$. 
    The shaded blue background region on the right of the source region is $\mathbf{BR_{out}}$ and is used to estimate the background spectrum in this analysis.
    The statistical similarity of the $\chi$-distributions in the two energy ranges substantiates our technique of scaling the background.
    (b) Same as in (a) expect for Compton scatter angle $\phi = [40^{\circ},41^{\circ}]$.
    } 
    \label{fig:chislices}
\end{figure}

The CDS background estimation routine is illustrated in Fig.~\ref{fig:spectrumslices} and Fig.~\ref{fig:chislices}. 
Figure~\ref{fig:spectrumslices} shows the flight $\mathbf{SR}$ spectrum in red for two different Compton scatter ranges: $\phi = [20^{\circ},21^{\circ}]$ and $\phi=[40^{\circ},41^{\circ}]$.
The difference seen in these two spectral shapes results from the $\phi$ energy-dependence of Compton scattering and clearly demonstrates the need to perform background estimation as a function of $\phi$.
The background spectrum from $\mathbf{BR}$ is plotted in blue after rescaling.
For each Compton scatter range, there is a very good match between the source region spectrum and the scaled background region spectrum, as shown in the residuals of the plots.

By using the energy range from 520--720~keV to scale the background $\phi$-dependent spectra, we are relying on the energy-independence of $\chi$. 
Figure~\ref{fig:chislices} shows the $\chi$ distribution from background simulations for two different energy ranges above and below the 511~keV line emission: 300--500~keV is shown in green and 520--720~keV is shown in black. 
Simulation data is used since the positron annihilation spectrum is known to have the o-Ps continuum below 511~keV and therefore the flight data should show statistical differences between the two energy intervals.
The $\chi$-distributions have been normalized, and we confirmed that there is no statistical difference between the two distributions with chi-square statistical tests. 

Overlaid on the $\chi$-histograms in Fig.~\ref{fig:chislices} are the locations of the source region and the background regions, highlighted in red and blue, respectively.
For each $\phi$ bin, we know the polar scatter angles that are consistent with the source region satisfy $|\chi - \phi| < \Delta$. 
The $\chi$ values that are consistent with the GC with a $\Delta=~$16$^{\circ}$ origin cut are shaded in red, while the $\chi$ values that are consistent with the two background regions are shaded in blue.
As discussed in \ref{sec:analysis}, we only use $\mathbf{BR_{out}}$ for this analysis, but the $\chi$ values that are consistent with $\mathbf{BR_{in}}$ is also shaded in this plot for clarification.
The total number of counts within the red $\mathbf{SR}$ of the 520--720~keV $\chi$-distribution in Fig.~\ref{fig:chislices}, by definition, is equal to the integrated spectrum of the $\mathbf{SR}$ within 520--720~keV in Fig.~\ref{fig:spectrumslices}. 
This is also true for the $\mathbf{BR}$.


\subsection{Spatial Background Estimation Routine}
\label{sec:spatialroutine}

We can determine the radial distribution of the emission by performing a CDS-ARM analysis. 
Analogous to the routine described in Sec.~\ref{sec:spectralroutine}, we want to define a CDS-ARM ``off-measurement'' to recover the angular distribution of the celestial signal.
To do this, we need to find an appropriate estimate of the background distribution.

Obtaining the CDS-ARM distribution of the Galactic positron annihilation emission is a direct measure of the radial extent of the source around the GC; if the emission originates from a point source, we would expect to recover a CDS-ARM distribution with a FWHM $\sim$6$^{\circ}$, equivalent to COSI's angular resolution, as shown in Figure~\ref{fig:2DCDSandARM}~(b).
Whereas if the emission has an inherent width, the measured CDS-ARM will be a convolution of the instrument point spread function and the spatial distribution of the source. 


The CDS-ARM background estimation procedure relies on the results from the spectral estimation and is a four-step process: 
\begin{enumerate}

\item{
Find a separate CDS-ARM distribution for the two different energy ranges: the line interval $(E | 506 < E < 516~\text{keV})$ and the higher-energy range $(E | 520 < E < 720~\text{keV})$. The $\phi$-dependent CDS-ARM distribution in the 511~keV line interval is given by $(\chi - \phi)$ for each $\phi$ in bin $i$: $N^{511}(\phi_i, \chi-\phi)$. This is our ``on-source'' measurement. The CDS-ARM distribution for the higher-energy (HE) interval is
$N^{HE}(\phi_i, \chi-\phi)$.
}

\item{Use the scaled background spectrum $B^{\text{SR}}(\phi, E)$ from the third step in the spectrum estimation routine to determine the normalization factor for the higher-energy CDS-ARM in each bin $i$:

\begin{equation}
A_{\phi_i} = \frac{B^{\text{SR}}(\phi_i,  E | 506 < E < 516)}{ B^{\text{SR}}(\phi_i,  E | 520 < E < 720)}.
\end{equation}
}

\item{For each $\phi$ bin $i$, scale the high-energy interval CDS-ARM distribution by $A_{\phi_i}$ to obtain an estimate for the background distribution,
\begin{equation}
B^{511}(\phi_i, \chi-\phi) = A_{\phi_i} N^{HE}(\phi_i, \chi-\phi),
\end{equation}
i.e., an estimate for an ``off-measurement.''
}

\item{
With our ``on-source'' measurements $N^{511}(\phi_i, \chi-\phi)$ from Step~1 and our background distribution estimate $B^{511}(\phi_i, \chi-\phi)$ from Step~3, we can find the radial distribution of the source by summing the remaining counts in each $\phi$ bin:
\begin{equation}
S(\chi-\phi) = \sum_i \left [ N^{511}(\phi_i, \chi-\phi) - B^{511}(\phi_i, \chi-\phi) \right ].
\end{equation}
}
\end{enumerate}

\begin{figure}[tp]
\centering
\subfloat[{CDS-ARM for $\phi= [16^{\circ},17^{\circ}]$.}]{\includegraphics[width=8.5cm]{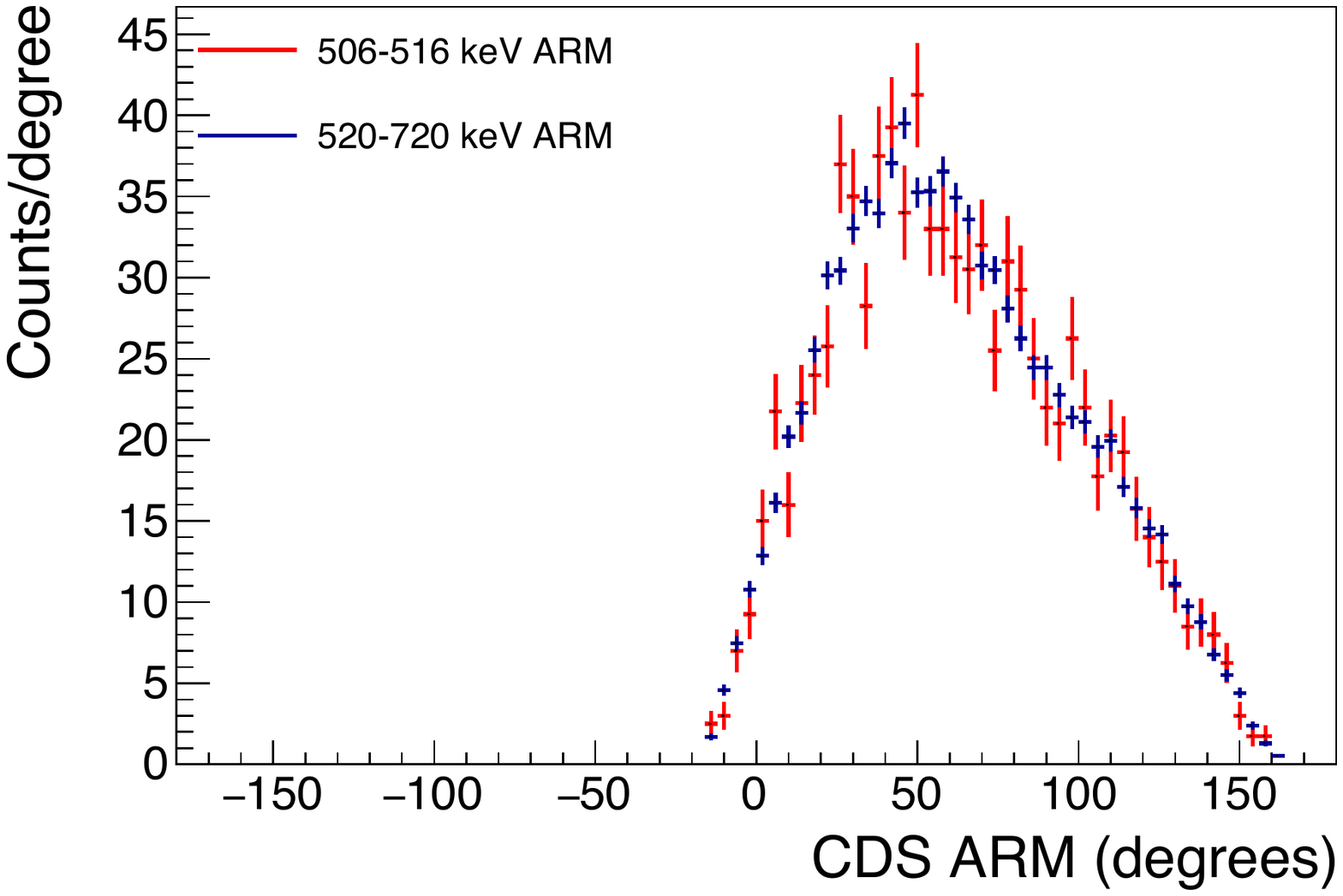}}\\
\subfloat[{CDS-ARM for $\phi = [59^{\circ},60^{\circ}]$.}]{\includegraphics[width=8.5cm]{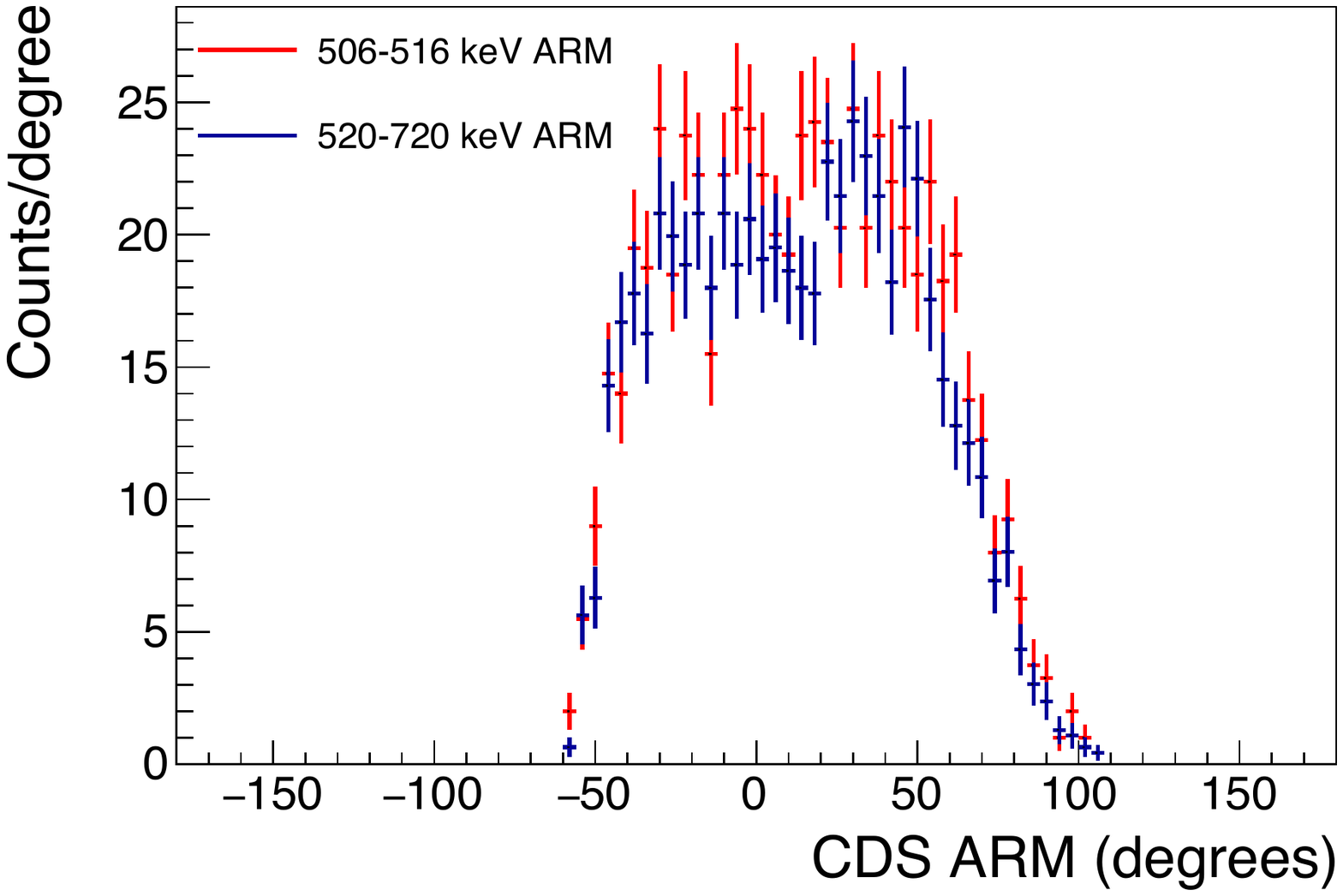}}
\caption{The CDS-ARM distribution $(\chi - \phi)$ around the GC from the background simulation for events with Compton scatter angles $[16^{\circ},17^{\circ}]$ and $[59^{\circ},60^{\circ}$], in (a) and (b), respectively. Events with $E_{\gamma} =~$506--516~keV are plotted in red. The events from the 520--720~keV interval have been scaled as described by Step~3 of the routine in Sec.~\ref{sec:spatialroutine}. 
In each histogram, the $x$-axis represents the radial distribution around the GC.
}
\label{fig:phidependentarmsub}
\end{figure}

Figure~\ref{fig:phidependentarmsub} shows the CDS-ARM distribution from a full flight background simulation (see \citet{kieransthesis2018} for details) for events with Compton scatter angle $\phi = [16^{\circ},17^{\circ}]$ shown in (a), and events with $\phi=[59^{\circ},60^{\circ}]$ shown in (b).
The CDS-ARM distribution for the line interval 506--516~keV is shown in red. 
The distribution from the higher-energy range, 520--720~keV, has been scaled using the background spectra $B^{\text{SR}}(\phi, E)$, and is shown in blue. 
This scaled higher-energy CDS-ARM serves as our estimated radial distribution of the background and closely matches the line interval distribution.

\subsection{Background Estimation Method Validation}


We developed a detailed background simulation for the full COSI flight, including the atmospheric contribution as well as instrument activation, that closely matches the measured data. 
With simulations of GC sources, we were able to recover the simulated flux, with the correct line width and spatial distribution.
See \citet{kieransthesis2018} for a detailed description of the method validation.

\section{Results}
\label{sec:results}

\subsection{Positron Annihilation Spectrum}
\label{sec:spectralresults}

\begin{figure}[t]
\centering
\includegraphics[width=9cm]{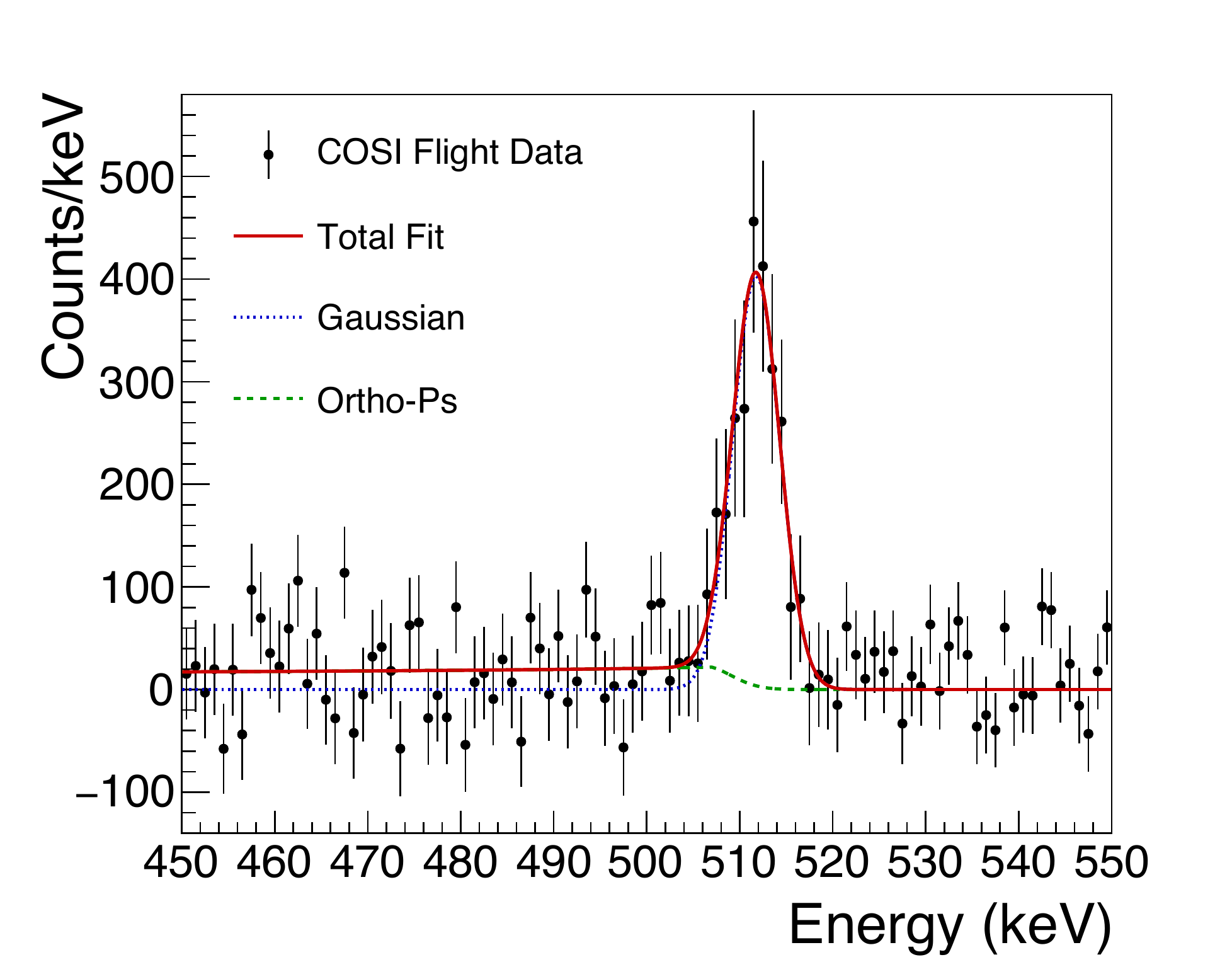}
\caption{Positron annihilation spectrum from the 2016 COSI flight from a 16$^{\circ}$ source region around the GC. 
The spectrum is fit with a single Gaussian component to describe the 511~keV line and the theoretical o-Ps continuum spectrum. The total number of counts in the 511~keV line is 2560$\pm$300~cts. The detection significance is 7.2$\sigma$ detection.}
\label{fig:511spectrum}
\end{figure}


Figure~\ref{fig:511spectrum} shows our final measured spectrum for a 16$^{\circ}$ origin cut around the GC after applying the CDS background estimation routine described in Sec.~\ref{sec:spectralroutine}. 
The significance of the Galactic 511~keV line is 7.2$\sigma$ \citep[calculated with an F-test;][]{ftest}. 
The event selections for this analysis are listed in Tab.~\ref{tb:511spectrumeventselections}.

\begin{table}[t]
\centering
\begin{tabular}[c]{lcc}
\hline
\multicolumn{2}{l}{Parameter}			& 	Value			    \\
\hline
Gaussian Fit	&	$\mu$				& 511.8$\pm$0.3~keV	    \\
		    	&	$\sigma$    		& 2.5$\pm$0.3~keV 		\\	
			    &	$A$					& 403$\pm$57~cts/keV	\\
o-Ps Fit	    & 	$B$					& 12$\pm$4~cts/keV		\\
\hline
$\chi^2$/d.o.f. &						& 193.0/196		    	\\
\hline
$I_{2\gamma}$   &                   	& 2560$\pm$300~cts		\\
$I_{3\gamma}$   &               		& 5110$\pm$1700~cts	    \\
\hline
$f_{Ps}$		&						& 0.76$\pm$0.12		    \\
\hline
\end{tabular}
\caption{Fit parameters for the COSI 2016 flight positron annihilation  spectrum shown in Figure~\ref{fig:511spectrum}. The fit is made over the energy range 450--550~keV. 
The reduced $\chi^2$ of 0.99 with 196 degrees of freedom (d.o.f.) implies an adequate fit to the spectrum. The derived parameters are the integrated 511~keV line counts, $I_{2\gamma}$, as well as the o-Ps counts ($I_{3\gamma}$; area under the curve) as well as the positronium fraction.}
\label{tb:511resultsfitpars}
\end{table}


As discussed in Section~\ref{sec:intro}, the positron annihilation emission from the Galaxy is characterized by two spectral signatures: the annihilation line at 511~keV and the o-Ps continuum below 511~keV. A possible contribution from the diffuse Galactic $\gamma$-ray continuum is strongly suppressed by this method. We therefore describe the spectrum by combining a Gaussian and o-Ps spectral component, $F_{oPs}(E)$ as defined in \citet{ore}, to give a 4-parameter spectral fit function:
\begin{equation}
\label{eq:spectralfit}
F(E) = A \exp\left(-\frac{(E-\mu)^2}{2\sigma^2}\right) + B F_{oPs}(E),
\end{equation}
where $A$, $B$, $\mu$ and $\sigma$ are the free parameters of the fit. $A$ and $B$ are amplitude scaling factors for each spectral component, and $\mu$ and $\sigma$ are the Gaussian mean and width, respectively.
$F_{oPs}(E)$ has been convolved with the COSI instrument response prior to the fit. 
From the relative flux of the o-Ps continuum and the 511~keV line, denoted by $I_{3\gamma}/I_{2\gamma}$, we calculate the positronium fraction~\citep{prantzos2011}:
\begin{equation}
f_{Ps} = \frac{8  I_{3\gamma}/I_{2\gamma}}{9 + 6  I_{3\gamma}/I_{2\gamma}}.
\end{equation}
The resulting fit parameters are listed in Tab.~\ref{tb:511resultsfitpars}.

The line was found with a centroid at 511.8$\pm$0.3~keV with a width of $\sigma=2.5\pm$0.3~keV.
The integrated background spectrum over the whole flight gives an annihilation line centroid of 511.54$\pm$0.01~keV, which defines a systematic offset in the energy calibration around these energies. 
The fitted Galactic 511~keV centroid is consistent with this calibrated value.
Accounting for the spectral resolution of COSI, 1.85$\pm$0.1~keV at 511~keV~\citep{kieransthesis2018}, we estimate the line broadening of the celestial positron annihilation line to be $\sigma=1.7\pm$0.4~keV. 
In comparison, \citet{siegert2019} report an average Galactic line width of 2.43$\pm$0.14~keV and centroid at 511.05$\pm$0.03~keV. 

In the spectral study by \citet{jean2006}, the authors reported a better fit to the Galactic 511~keV line with two Gaussian components, a narrow line with $\sigma=~$0.6~keV and a broad line with $\sigma=~$2.3~keV; however, the fit to the COSI data, which has a $\chi^2$/d.o.f. of 193.0/196, is satisfactory with only a single Gaussian. 

The relative intensity of the o-Ps continuum, which has a 3.0$\sigma$ detection significance compared to the null hypothesis, 
results in a surprisingly low $f_{Ps}$ of 0.76$\pm$0.12. 
This is a smaller fraction than other reported measurements. 
For example, from analysis of SPI data, \citet{siegert2016a} report an o-Ps fraction of $1.080\pm 0.029$, and \citet{jean2006} find $f_{Ps} = 0.97\pm0.02$. 
Measurements with other instruments are also consistent with $f_{Ps}  \sim~$1, where \citet{kinzer1996} find $f_{Ps} = 0.97\pm0.03$ with OSSE, and \citet{harris1998} reported $0.94\pm0.4$ from TGRS. This discrepancy is further discussed in Sec.~\ref{sec:discussion}.

\subsection{Measured Flux}
\label{sec:fluxresults}


To convert the measured counts into a source flux, we use simulations to estimate the effective area, $A_{Eff}$, of COSI at 511~keV.
We simulate a Galactic 511~keV source based on the Skinner Model \citep{skinner2014}, with 10 times the expected Galactic flux to increase statistics. 
For these simulations, we use the COSI flight aspect information and take into account the drops in altitude to calculate the correct exposure and attenuation in the atmosphere.
The flux is then calculated to be
\begin{equation}
\label{eq:effcalcskinner}
\begin{split}
\text{Flux} & = \frac{N_{det}}{A_{Eff} \times \text{time}} \\
& = \frac{2560 \pm 300~\text{cts}}{\left( \frac{8775 \text{cts}}{0.0133 \mathrm{\gamma\,cm^{-2}\,s^{-1}} \times 3.08 \times 10^6 \text{s}} \right) \times 3.08\times10^6~\text{s}} \\
& =  (3.9\pm 0.4) \times 10^{-3} ~\mathrm{\gamma\,cm^{-2}\,s^{-1}}.
\end{split}
\end{equation}
The exposure time from the full flight is 3.08$\times$10$^6$~s, ignoring times of very low altitude. From full flight simulations we find 8775~cts between 506--516~keV from the 16$^{\circ}$ region around the GC, assuming a flux of 0.0133~$\mathrm{\gamma\,cm^{-2}\,s^{-1}}$. These numbers allow us to calculate an effective area, written out above with the numbers from the simulation, to find the measured flux of $(3.9\pm 0.4) \times 10^{-3}~\mathrm{\gamma\,cm^{-2}\,s^{-1}}$ from the COSI observations. The error is statistical and does not include all systematics.
For comparison, \citet{siegert2016a} report a total Galactic 511~keV line flux of (2.74$\pm$0.03)$\times$10$^{-3}~\mathrm{\gamma\,cm^{-2}\,s^{-1}}$ with SPI measurements, and \citet{purcell1997} find a line flux of $(2.25\pm0.07)\times 10^{-3}~\mathrm{\gamma\,cm^{-2}\,s^{-1}}$ by combining OSSE, SMM, and TGRS data.

\subsection{Radial Distribution of the 511 keV Sky}
\label{sec:spatialresults}

\begin{figure}[t]
\centering
\includegraphics[width = 9cm]{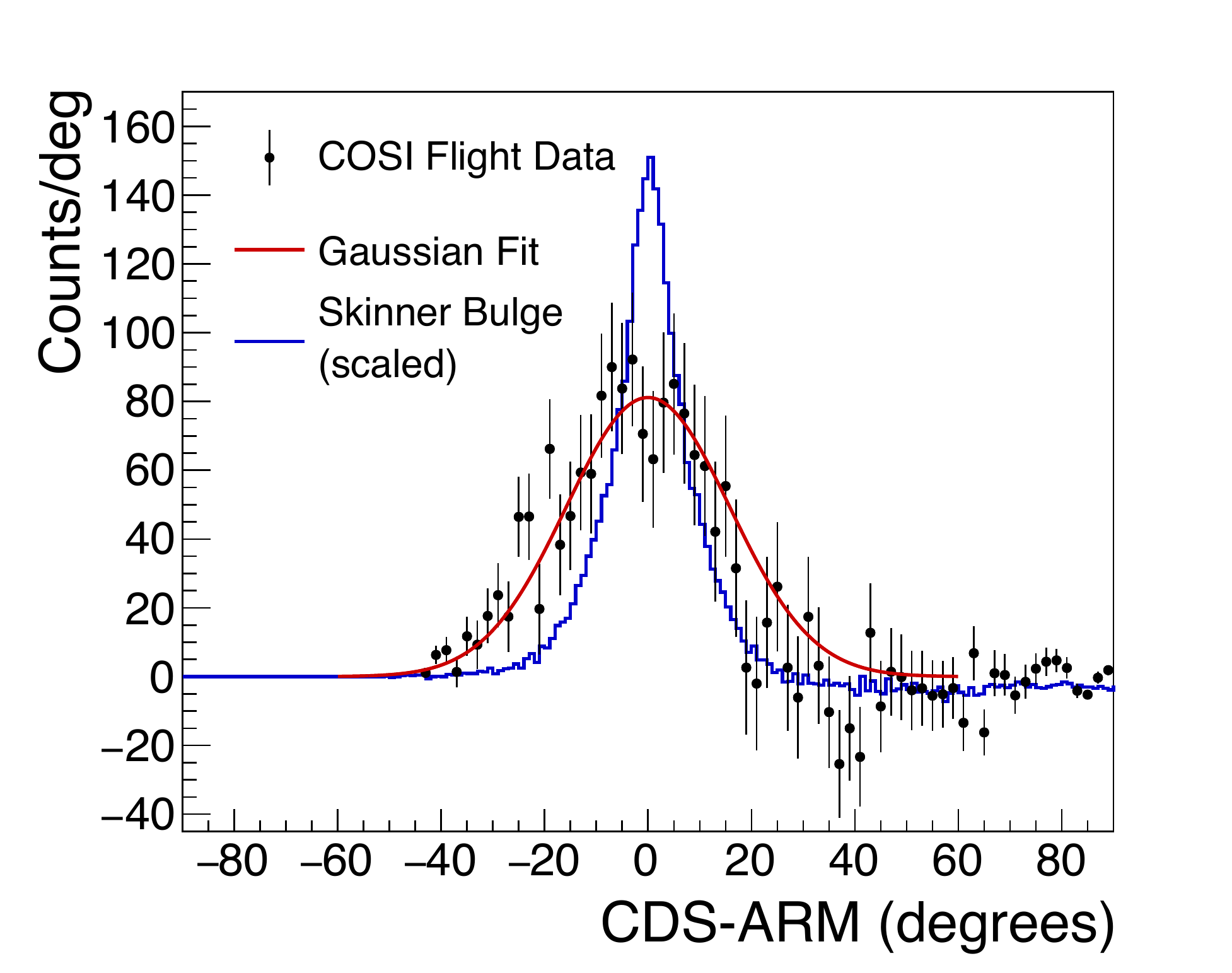}
\caption{Measured radial distribution of emission around the GC} from the 2016
COSI flight CDS-ARM for $E = 506$--516~keV. The distribution is fit with a single Gaussian, which gives a FWHM of $33\pm2^{\circ}$. The event selections for this
analysis are listed in Tab.~\ref{tb:511spectrumeventselections}, and the best-fit parameters are listed in Tab.~\ref{tb:511resultsarmfitpars}.
\label{fig:511ARM}
\end{figure}

COSI provides a novel investigation of the 511~keV spatial distribution. 
As discussed in Sec.~\ref{sec:intro}, coded mask telescopes, like SPI, rely on a model fitting approach to determine the morphology of the Galactic positron annihilation emission and suffer from an inability to detect diffuse emission with strong gradients, and thus have limitations. 
Furthermore, collimated instruments, such as OSSE, use on/off pointings with a detector that has no inherent imaging capabilities.
Not only is the imaging model dependent, but these telescope types favor certain angular scales (e.g. FOV of collimator, FOV and pitch of coded mask).
A Compton telescope, like COSI, does not favor a particular spatial frequency and allows for a more direct measurement of the extent of the Galactic emission because Compton telescopes obtain spatial information from every photon.

Figure~\ref{fig:511ARM} shows the measured radial distribution for the 511~keV line with a 40$^{\circ}$ pointing selection on the GC.
This CDS-ARM distribution is fit with a single Gaussian and the measured FWHM of 33$\pm$2$^{\circ}$. 
The parameters from the Gaussian fit of the CDS-ARM distribution are shown in Tab.~\ref{tb:511resultsarmfitpars}. 
A 40$^{\circ}$ pointing selection was used for this analysis because it was found to decrease the uncertainties of the measured distribution; when no pointing selection is used, the measured FWHM is 32$\pm$4$^{\circ}$.

As described in Sec.~\ref{sec:analysis}, the CDS-ARM distribution is the angular distance of each 511~keV Compton event from the GC. Therefore, this 1-D distribution shows the radial extent of emission around the GC. Unfortunately, it is not able to separate out any difference in longitude and latitude or possible asymmetries. However, work is currently underway to produce a full sky image from the COSI 2016 flight data (Siegert et al. in prep).

We note that through detailed simulations, we have concluded that we are only sensitive to the bulge emission with the current data set and techniques and are not able to detect a disk component; therefore, we compare our measured distribution with the simulated CDS-ARM distribution of the Skinner Model bulge emission only, shown in blue. 
The COSI data shows a distribution that is significantly larger.
Likewise, the measured distribution from combined OSSE/SMM/TGRS data has also been found to be a much narrower profile around the GC that the radial distribution reported here~\citep{purcell1997,kinzer2001}.

\begin{table}[t]
\centering
\begin{tabular}[c]{lcc}
\hline
\multicolumn{2}{l}{Parameter}		& Value				    \\
\hline
Gaussian Fit	&	$\mu$			& fixed at 0			\\
				&	$\sigma$		& 14.0$\pm$0.7$^{\circ}$\\	
				&	$A$				& 89$\pm$0.6~cts/deg		\\
\hline
$\chi^2$/d.o.f. &	    			& 52.1/52				\\
\hline
FWHM    		&					& 33$\pm$2$^{\circ}$		\\
\hline
\end{tabular}
\caption{Fit parameters for the flight data CDS-ARM distribution shown in Figure~\ref{fig:511ARM}. The distribution is fit with a single Gaussian since the statistics are insufficient to warrant more parameters.}
\label{tb:511resultsarmfitpars}
\end{table}

The Skinner Model bulge distribution in Fig.~\ref{fig:511ARM} has been scaled so that the area under the curve is the same as the flight data CDS-ARM distribution. 
This visually shows the difference in widths of the Skinner Model bulge distribution and the detected spatial distribution at 511~keV.
To test the difference between these two histograms, we perform a chi-squared test, which gives a P-value~=~0.001, and therefore there is a 3$\sigma$ statistical significance between the COSI distribution and the Skinner bulge distribution. 

The radial distributions shown in Fig.~\ref{fig:511ARM} include the COSI instrument response, which can be subtracted-off in quadrature to determine the true emission around the GC. With the inherent COSI angular resolution of 6$^{\circ}$ at 511~keV being small relative to the width of the measured CDS-ARM distribution, the reported width of the 511~keV Galactic emission is $32\pm2^{\circ}$.

\section{Discussion}
\label{sec:discussion}

The spectral results from the 2016 COSI flight show a measured 511~keV line at 511.8$\pm$0.3~keV with $\sigma=1.7\pm$0.4~keV from a 16$^{\circ}$ region around the GC. This corresponds to a measured flux of $(3.9\pm0.4)\times10^{-3}$~$\mathrm{\gamma\,cm^{-2}\,s^{-1}}$.
We measure the o-Ps continuum emission with $5110\pm1700$ total counts in the distribution, which corresponds to a 3$\sigma$ detection. From the ratio of the 511~keV line and the o-Ps flux, we find a positronium fraction of $f_{Ps}=0.76\pm0.12$. These line flux and o-Ps continuum are within $\sim3\sigma$ of previously reported values from SPI measurements.

We find a slightly enhanced 511~keV flux which is 1.4 times larger than the total Galactic flux reported in \citet{siegert2016a} from SPI data and 1.7 times the flux reported for combined OSSE/SMM/TGRS observations~\citep{purcell1997}. Ever since the first measurements of the 511~keV emission in the 1970's, determining the true flux has proven to be a challenge with wider FOV instruments always recording a larger flux due to the diffuse nature of the source (see Figure~4 of \citet{purcell1997}).
Furthermore, the flux results from SPI and OSSE/SMM/TGRS have relied on an assumed spatial model.
Nonetheless, we must consider the systematics which could result in an overestimated flux, particularly since the backgrounds at these energies are known to be heavily influenced by even small variations of the balloon environment.
We performed detailed background simulations and a thorough validation of our analysis method for Galactic source models (see \citet{kieransthesis2018} for details). 
The simulation results substantiate that we can determine the correct spectral shapes and that the $f_{Ps}$ is preserved through the background estimation technique described here with simulated data.

\begin{figure}
\centering
\includegraphics[width=9cm]{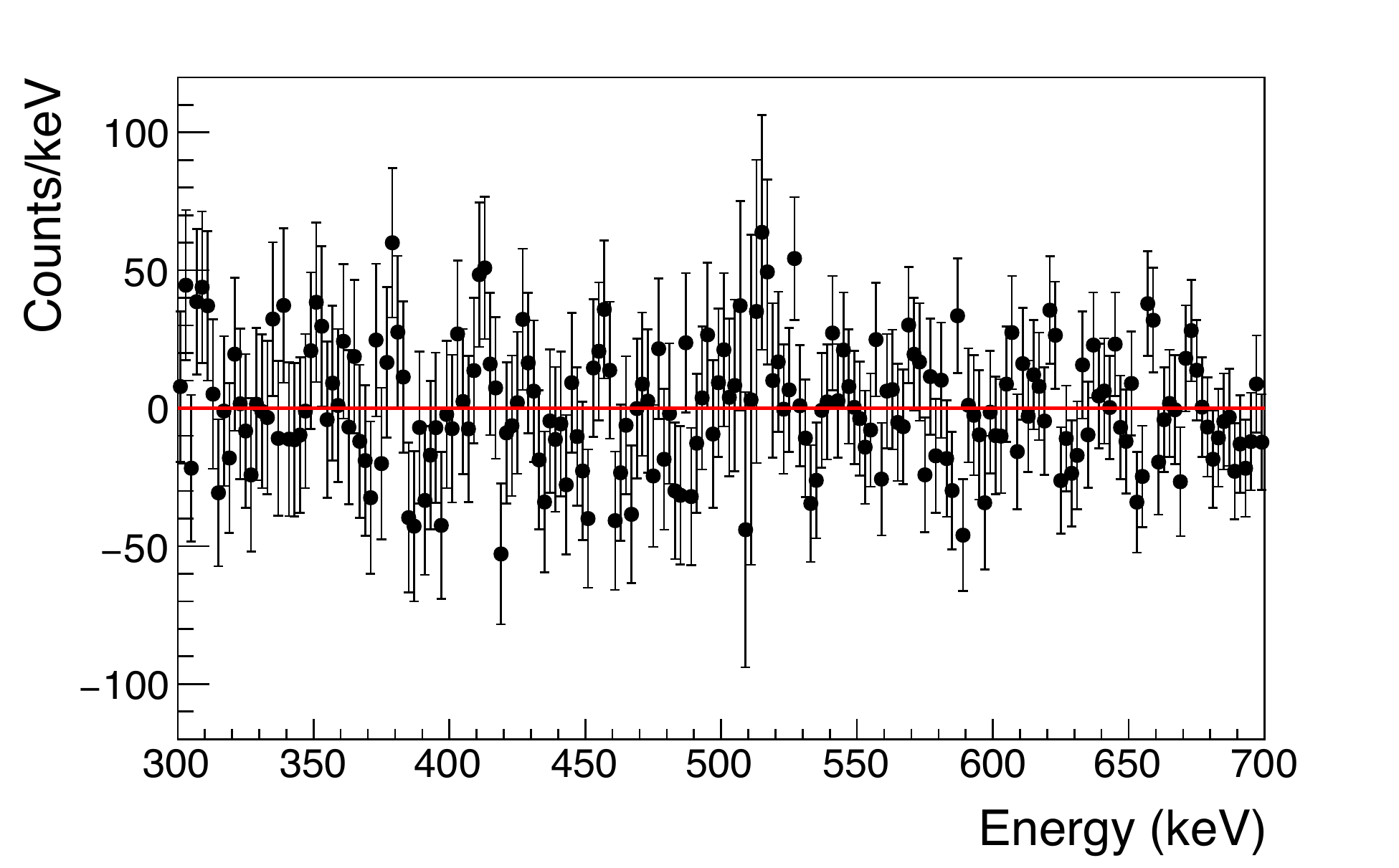}
\caption{CDS background-subtracted spectrum of flight data with Galactic coordinates $(120^{\circ}, -60^{\circ})$ chosen to be the source location. The flat spectrum further confirms that the systematics in the CDS subtraction are minor.}
\label{fig:offaxisflightsub}
\end{figure}

We test our method at different origin cuts in the sky which we expect to be void of positron annihilation as a further check.
Figure~\ref{fig:offaxisflightsub} shows the resulting spectrum when the source location is chosen to be at $(120^{\circ}, -60^{\circ})$ in Galactic coordinates.
The nearly flat spectrum further confirms the legitimacy of our routine. 

Although systematics are known to be high for Galactic 511~keV measurements, it is useful to consider what these spectral results could imply. 
One possible explanation for a higher flux is that the true spatial distribution does not agree with the distribution modeled for the SPI observations. 
For example, a larger disk contribution in the measurement would result in a larger number of events that are consistent with the inner region of the Galaxy, and therefore our simulation of the flight-averaged effective area using the Skinner Model (Sec.~\ref{sec:fluxresults}) would result in a falsely high flux. However, this scenario does not offer an explanation for the low $f_{Ps}$.

The unusually low reported value of $f_{Ps}$, though only a 3$\sigma$ significance, could be a signature of a previously-undetected emission component with a smaller $f_{Ps}$. 
From the measured line shape and a positronium fraction close to 1, analysis of SPI data has concluded that the annihilation is occurring predominately in warm and neutral phases of the ISM~\citep{jean2006, churazov2005}.
An $f_{Ps}<1$ could be due to annihilation in a dusty warm phase of the ISM, which predicts a narrow 511~keV line and a suppressed $f_{Ps}$~\citep{zurek1985, guessoum2005}.
However, with our statistics and systematics, further investigation is necessary before any conclusions can be reached from these measurements.

The measured radial distribution of the 511~keV line around the GC shows a Gaussian shape with a FWHM of $33\pm2^{\circ}$ (convolved with the instrument response).
The proposed models from \citet{skinner2014} and \citet{siegert2016a} have a Galactic bulge distribution defined by three components (Sec.~\ref{sec:intro}), and when convolved with the COSI instrument response, the width of this distribution is 15.0$^{\circ}$; therefore, we measure a distribution that is twice as broad.

The large extended shape of the COSI detected CDS-ARM distribution is intriguing. 
\citet{siegert2016a} use the same bulge description as \citet{skinner2014} for their spectral studies, but report an alternate, yet equally significant, model in the paper's appendix. 
This alternate model describes the bulge emission by two elongated components, with a longitudinal extent of the broad bulge up to FWHM~$\sim$~55$^{\circ}$. 
\citet{bouchet2010} also report a halo morphology that is consistent with SPI data; with this extended component, the 511~keV flux was found to be $2.9\times10^{-3}~\mathrm{\gamma\,cm^{-2}\,s^{-1}}$, which is closer to the flux reported here.
Furthermore, \citet{skinner2012} performed analysis with 10~years of SPI data combined with archival data from OSSE, SMM, and TGRS to conclude that the absence of an extended halo would be inconsistent with the OSSE/SMM/TGRS dataset.
The COSI results seem to agree more with the halo morphology; however, more data is needed to make a strong conclusion about the spatial distribution.

A connection between the potentially overestimated flux in the spectral estimation and the broad spatial distribution must also be considered.
These measurements are indeed related, since the background spectral subtraction determines the background CDS-ARM distribution scaling factors. 
If the excess counts in the detected 511~keV line are from a poorly modeled background component, then it is most likely that the background component would have a spatial distribution similar to the FOV of the instrument; however, this would give a FWHM~$>50^{\circ}$, which is much larger than the distribution that we measure and therefore seems improbable.

\vspace{-0.2cm}
\section{Conclusions}
\label{sec:conclusions}

We have reported the first detection of the positron annihilation signal from the Galaxy with a compact Compton telescope, and to perform this analysis, we have developed an accurate background estimation technique that is valid for sources of line emission. 

We have found a $7.2\sigma$ detection of the 511~keV line from the Galactic center region, and a $3.0\sigma$ detection of the o-Ps continuum after 46~days of flight. 
The relative ratio of the 511~keV line and o-Ps continuum results in a low measurement of $f_{Ps} = 0.76\pm$0.12. Through this analysis, we found the radial distribution of the 511~keV emission around the Galactic center to be described by a Gaussian with FWHM of $32\pm2^{\circ}$. 

Although our analysis techniques are still being improved, the results from this study of Galactic positron annihilation are intriguing. 
The flux measurements, although with known systematics, could hint at a morphology that is not seen in SPI observations. The measured angular distribution is broader than the emission models presented by the SPI collaboration and is similar to what is expected from a halo model. 
The results discussed here show the power of Compton telescopes and the CDS analysis, and the need for more data is clear.



\acknowledgments
{The authors thank the NASA Columbia Scientific Ballooning Facility team for their support of a successful COSI flight. Special thanks goes to Brent Mochizuki and Steve McBride for their engineering support of the COSI instrument.

Support for COSI is provided by NASA Astrophysics Research and Analysis grant NNX14AC81G. This work is also supported in part by CNES. C.~Kierans is supported by a NASA Postdoctoral Program Fellowship. C.~Sleator is supported by a NASA Earth and Space Sciences Fellowship. T.~Siegert is supported by the German Research Society (DFG-Forschungsstipendium SI 2502/1-1).}


\bibliographystyle{yahapj}
\bibliography{references}

\begin{thebibliography}{}
\providecommand\natexlab[1]{#1}
\providecommand\JournalTitle[1]{#1}

\bibitem[{Agostinelli {et~al.}(2003)Agostinelli, Allison, Amako, Apostolakis,
  Araujo, Arce, Asai, Axen, Banerjee, Barrand, Behner, Bellagamba, Boudreau,
  Broglia, Brunengo, Burkhardt, Chauvie, Chuma, Chytracek, Cooperman, Cosmo,
  Degtyarenko, Dell'acqua, Depaola, Dietrich, Enami, Feliciello, Ferguson,
  Fesefeldt, Folger, Foppiano, Forti, Garelli, Giani, Giannitrapani, Gibin,
  Gomez, Gonzalez, Gracia, Greeniaus, Greiner, Grichine, Grossheim, Guatelli,
  Gumplinger, Hamatsu, Hashimoto, Hasui, Heikkinen, Howard, Ivanchenko,
  Johnson, Jones, Kallenbach, Kanaya, Kawabata, Kawabata, Kawaguti, Kelner,
  Kent, Kimura, Kodama, Kokoulin, Kossov, Kurashige, Lamanna, Lampen, Lara,
  Lefebure, Lei, Liendl, Lockman, Longo, Magni, Maire, Medernach, Minamimoto,
  Mora, Morita, Murakami, Nagamatu, Nartallo, Nieminen, Nishimura, Ohtsubo,
  Okamura, O'Neale, Oohata, Paech, Perl, Pfeiffer, Pia, Ranjard, Rybin,
  Sadilov, Di~Salvo, Santin, Sasaki, Savvas, Sawada, Scherer, Sei, Sirotenko,
  Smith, Starkov, Stoecker, Sulkimo, Takahata, Tanaka, Tcherniaev, Safai,
  Tropeano, Truscott, Uno, Urban, Urban, Verderi, Walkden, Wander, Weber,
  Wellisch, Wenaus, Williams, Wright, Yamada, Yoshida, \& Zschiesche}]{geant4}
Agostinelli, S., Allison, J., Amako, K., {et~al.} 2003,
  \href{http://dx.doi.org/https://doi.org/10.1016/S0168-9002(03)01368-8}{\JournalTitle{\nima},
  506, 250}

\bibitem[{{Alexis} {et~al.}(2014){Alexis}, {Jean}, {Martin}, \&
  {Ferri{\`e}re}}]{alexis2014}
{Alexis}, A., {Jean}, P., {Martin}, P., \& {Ferri{\`e}re}, K. 2014,
  \href{http://dx.doi.org/https://doi.org/10.1051/0004-6361/201322393}{\JournalTitle{\aap},
  564, A108}

\bibitem[{Amman {et~al.}(2007)Amman, Luke, \& Boggs}]{amman2007}
Amman, M., Luke, P.~N., \& Boggs, S.~E. 2007,
  \href{http://dx.doi.org/https://doi.org/10.1016/j.nima.2007.05.307}{\JournalTitle{\nima},
  579, 886}

\bibitem[{Bandstra {et~al.}(2011)Bandstra, Bellm, Boggs, Perez-Berker,
  Zoglauer, Chang, Chiu, Liang, Chang, \& Liu}]{bandstra2011}
Bandstra, M.~S., Bellm, E.~C., Boggs, S.~E., {et~al.} 2011,
  \href{http://dx.doi.org/https://doi.org/10.1088/0004-637X/738/1/8}{\JournalTitle{\apj},
  738, 8}

\bibitem[{Bartels {et~al.}(2018)Bartels, Calore, Storm, \&
  Weniger}]{Bartels2018_binaries511}
Bartels, R., Calore, F., Storm, E., \& Weniger, C. 2018,
  \href{http://dx.doi.org/10.1093/mnras/sty2135}{\JournalTitle{\mnras}, 480,
  3826}

\bibitem[{Beacom \& Y\"{u}ksel(2006)}]{beacom2006}
Beacom, J.~F., \& Y\"{u}ksel, H. 2006,
  \href{http://dx.doi.org/https://doi.org/10.1103/PhysRevLett.97.071102}{\JournalTitle{\prl},
  97, 071102}

\bibitem[{B{\oe}hm \& Ascasibar(2004)}]{boehm2004b}
B{\oe}hm, C., \& Ascasibar, Y. 2004,
  \href{http://dx.doi.org/https://doi.org/10.1103/PhysRevD.70.115013}{\JournalTitle{\prd},
  70, 115013}

\bibitem[{Boggs \& Jean(2000)}]{boggs2000}
Boggs, S.~E., \& Jean, P. 2000,
  \href{http://dx.doi.org/https://doi.org/10.1051/aas:2000107}{\JournalTitle{\aaps},
  145, 311}

\bibitem[{{Bouchet} {et~al.}(2010){Bouchet}, {Roques}, \&
  {Jourdain}}]{bouchet2010}
{Bouchet}, L., {Roques}, J.~P., \& {Jourdain}, E. 2010,
  \href{http://dx.doi.org/10.1088/0004-637X/720/2/1772}{\JournalTitle{\apj},
  720, 1772}

\bibitem[{Churazov {et~al.}(2005)Churazov, Sunyaev, Sazonov, Revnitsev, \&
  Varshalovich}]{churazov2005}
Churazov, E., Sunyaev, R., Sazonov, S., Revnitsev, M., \& Varshalovich, D.
  2005,
  \href{http://dx.doi.org/https://doi.org/10.1111/j.1365-2966.2005.08757.x}{\JournalTitle{\mnras},
  357, 1377}

\bibitem[{Crocker {et~al.}(2017)Crocker, Ruiter, Seitenzahl, Panther, Sim,
  Baumgardt, M{\"o}ller, Nataf, Ferrario, Eldridge, White, Tucker, \&
  Aharonian}]{crocker2017}
Crocker, R.~M., Ruiter, A.~J., Seitenzahl, I.~R., {et~al.} 2017,
  \href{http://dx.doi.org/https://doi.org/10.1038/s41550-017-0135}{\JournalTitle{\nata},
  1, 0135}

\bibitem[{Dermer \& Skibo(1997)}]{dermer1997}
Dermer, C.~D., \& Skibo, J.~G. 1997,
  \href{http://stacks.iop.org/1538-4357/487/i=1/a=L57}{\JournalTitle{The
  Astrophysical Journal Letters}, 487, L57}

\bibitem[{Deutsch(1951)}]{deutsch1951}
Deutsch, M. 1951,
  \href{http://dx.doi.org/https://doi.org/10.1103/PhysRev.82.455}{\JournalTitle{\physr},
  82, 455}

\bibitem[{{Guessoum} {et~al.}(2005){Guessoum}, {Jean}, \&
  {Gillard}}]{guessoum2005}
{Guessoum}, N., {Jean}, P., \& {Gillard}, W. 2005,
  \href{http://dx.doi.org/10.1051/0004-6361:20042454}{\JournalTitle{\aap}, 436,
  171}

\bibitem[{{Harris} {et~al.}(1998){Harris}, {Teegarden}, {Cline}, {Gehrels},
  {Palmer}, {Ramaty}, \& {Seifert}}]{harris1998}
{Harris}, M.~J., {Teegarden}, B.~J., {Cline}, T.~L., {et~al.} 1998,
  \href{http://dx.doi.org/10.1086/311429}{\JournalTitle{\apjl}, 501, L55}

\bibitem[{Haymes {et~al.}(1975)Haymes, Walraven, Meegan, Hall, Djuth, \&
  Shelton}]{haymes1975}
Haymes, R.~C., Walraven, G.~D., Meegan, C.~A., {et~al.} 1975,
  \href{http://dx.doi.org/https://doi.org/10.1086/153925}{\JournalTitle{\apj},
  201, 593}

\bibitem[{Higdon {et~al.}(2009)Higdon, Lingenfelter, \&
  Rothschild}]{higdon2009}
Higdon, J.~C., Lingenfelter, R.~E., \& Rothschild, R.~E. 2009,
  \href{http://dx.doi.org/https://doi.org/10.1088/0004-637X/698/1/350}{\JournalTitle{\apj},
  698, 350}

\bibitem[{Jean {et~al.}(2009)Jean, Gillard, Marcowith, \&
  Ferri\`{e}re}]{jean2009}
Jean, P., Gillard, W., Marcowith, A., \& Ferri\`{e}re, K. 2009,
  \href{http://dx.doi.org/10.1051/0004-6361/200809830}{\JournalTitle{\aap},
  508, 1099}

\bibitem[{Jean {et~al.}(2006)Jean, Kn{\"o}dlseder, Gillard, Guessoum,
  Ferri{\`e}re, Marcowith, Lonjou, \& Roques}]{jean2006}
Jean, P., Kn{\"o}dlseder, J., Gillard, W., {et~al.} 2006,
  \href{http://dx.doi.org/https://doi.org/10.1051/0004-6361:20053765}{\JournalTitle{\aap},
  445, 579}

\bibitem[{{Johnson III} {et~al.}(1972){Johnson III}, {Harnden Jr.}, \&
  Haymes}]{johnson1972}
{Johnson III}, W.~N., {Harnden Jr.}, F.~R., \& Haymes, R.~C. 1972,
  \href{http://dx.doi.org/https://doi.org/10.1086/180878}{\JournalTitle{\apj},
  172, L1}

\bibitem[{Kierans(2018)}]{kieransthesis2018}
Kierans, C.~A. 2018, PhD thesis, University of California, Berkeley

\bibitem[{{Kierans} {et~al.}(2016){Kierans}, {Boggs}, {Chiu}, {Lowell},
  {Sleator}, {Tomsick}, {Zoglauer}, {Amman}, {Chang}, {Tseng}, {Yang}, {Lin},
  {Jean}, \& {von Ballmoos}}]{kierans2016}
{Kierans}, C.~A., {Boggs}, S.~E., {Chiu}, J.-L., {et~al.} 2016,
  \href{https://arxiv.org/abs/1701.05558}{in Gamma-Ray Astrophysics in
  Multi-Wavelength Perspective, Proceedings in 11th INTEGRAL Conference}

\bibitem[{{Kinzer} {et~al.}(2001){Kinzer}, {Milne}, {Kurfess}, {Strickman},
  {Johnson}, \& {Purcell}}]{kinzer2001}
{Kinzer}, R.~L., {Milne}, P.~A., {Kurfess}, J.~D., {et~al.} 2001,
  \href{http://dx.doi.org/10.1086/322371}{\JournalTitle{\apj}, 559, 282}

\bibitem[{{Kinzer} {et~al.}(1996){Kinzer}, {Purcell}, {Johnson}, {Kurfess},
  {Jung}, \& {Skibo}}]{kinzer1996}
{Kinzer}, R.~L., {Purcell}, W.~R., {Johnson}, W.~N., {et~al.} 1996,
  \JournalTitle{\aaps}, 120, 317

\bibitem[{Klein \& Nishina(1929)}]{nishina}
Klein, O., \& Nishina, Y. 1929,
  \href{http://dx.doi.org/https://doi.org/10.1007/BF01366453}{\JournalTitle{\zap},
  52, 853}

\bibitem[{Kn{\"o}dlseder {et~al.}(1996)Kn{\"o}dlseder, von Ballmoos, Diehl,
  Oberlack, Sch{\"o}nfelder, Bloemen, Hermsen, Ryan, \&
  Bennett}]{knodlseder1996}
Kn{\"o}dlseder, J., von Ballmoos, P., Diehl, R., {et~al.} 1996,
  \href{http://dx.doi.org/https://doi.org/10.1117/12.254000}{in Proceedings of
  the SPIE, Vol. 2806, Gamma-Ray and Cosmic-Ray Detectors, Techniques, and
  Missions}

\bibitem[{Kn{\"o}dlseder {et~al.}(2005)Kn{\"o}dlseder, Jean, Lonjou,
  Weidenspointner, Guessoum, Gillard, Skinner, von Ballmoos, Vedrenne, Roques,
  Schanne, Teegarden, Sch{\"o}nfelder, \& Winkler}]{knodlseder2005}
Kn{\"o}dlseder, J., Jean, P., Lonjou, V., {et~al.} 2005,
  \href{http://dx.doi.org/10.1051/0004-6361:20042063}{\JournalTitle{\aap}, 441,
  513}

\bibitem[{Leventhal {et~al.}(1978)Leventhal, MacCallum, \&
  Stang}]{leventhal1978}
Leventhal, M., MacCallum, C.~J., \& Stang, P.~D. 1978,
  \href{http://dx.doi.org/https://doi.org/10.1086/182782}{\JournalTitle{\apj},
  225, L11}

\bibitem[{Ling {et~al.}(1977)Ling, Mahoney, Willett, \& Jacobson}]{ling1977}
Ling, J.~C., Mahoney, W.~A., Willett, J.~B., \& Jacobson, A.~S. 1977,
  \href{http://dx.doi.org/https://doi.org/10.1029/JA082i010p01463}{\JournalTitle{Journal
  of Geophysical Research}, 82, 1463}

\bibitem[{Lowell {et~al.}(2016)Lowell, Boggs, Chiu, Kierans, McBride, Tseng,
  Zoglauer, Amman, Chang, Jean, Lin, Sleator, Tomsick, von Ballmoos, \&
  Yang}]{lowell2016}
Lowell, A.~W., Boggs, S.~E., Chiu, J.~L., {et~al.} 2016,
  \href{http://dx.doi.org/https://doi.org/10.1117/12.2233145}{\JournalTitle{Proceedings
  of SPIE}, 9915}

\bibitem[{Lowell {et~al.}(2017{\natexlab{a}})Lowell, Boggs, Chiu, Kierans,
  Sleator, Tomsick, Zoglauer, Chang, Tseng, Yang, Jean, von Ballmoos, Lin, \&
  Amman}]{lowell2017b}
Lowell, A.~W., Boggs, S.~E., Chiu, C.~L., {et~al.} 2017{\natexlab{a}},
  \href{http://dx.doi.org/https://doi.org/10.3847/1538-4357/aa8ccd}{\JournalTitle{\apj},
  848, 120}

\bibitem[{Lowell {et~al.}(2017{\natexlab{b}})Lowell, Boggs, Chiu, Kierans,
  Sleator, Tomsick, Zoglauer, Chang, Tseng, Yang, Jean, von Ballmoos, Lin, \&
  Amman}]{lowell2017a}
---. 2017{\natexlab{b}},
  \href{http://dx.doi.org/https://doi.org/10.3847/1538-4357/aa8ccb}{\JournalTitle{\apj},
  848, 119}

\bibitem[{Millan \& Thorne(2007)}]{millan2007}
Millan, R.~M., \& Thorne, R.~M. 2007,
  \href{http://dx.doi.org/https://doi.org/10.1016/j.jastp.2006.06.019}{\JournalTitle{\jastp},
  69, 362}

\bibitem[{Milne {et~al.}(2001)Milne, Kurfess, Kinzer, \& Leising}]{milne2001}
Milne, P.~A., Kurfess, J.~D., Kinzer, R.~L., \& Leising, M.~D. 2001,
  \JournalTitle{AIP Conference Proceedings}, 587, 11

\bibitem[{Milne {et~al.}(1999)Milne, The, \& Leising}]{milne1999}
Milne, P.~A., The, L.-S., \& Leising, M.~D. 1999,
  \href{http://dx.doi.org/https://doi.org/10.1086/313262}{\JournalTitle{\apjs},
  124, 503}

\bibitem[{Mohorovicic(1934)}]{mohorovici1934}
Mohorovicic, S. 1934,
  \href{http://dx.doi.org/https://doi.org/10.1002/asna.19342530402}{\JournalTitle{\an},
  253, 93}

\bibitem[{Ore \& Powell(1949)}]{ore}
Ore, A., \& Powell, J.~L. 1949,
  \href{http://dx.doi.org/https://doi.org/10.1103/PhysRev.75.1696}{\JournalTitle{\physr},
  75, 1696}

\bibitem[{Prantzos(2006)}]{prantzos2006}
Prantzos, N. 2006,
  \href{http://dx.doi.org/https://doi.org/10.1051/0004-6361:20052811}{\JournalTitle{\aap},
  449, 869}

\bibitem[{Prantzos {et~al.}(2011)Prantzos, Boehm, Bykov, Diehl, Ferri{\`e}re,
  Guessoum, Jean, Kn{\"o}dlseder, Marcowith, \& Moskalenko}]{prantzos2011}
Prantzos, N., Boehm, C., Bykov, A.~M., {et~al.} 2011,
  \href{http://dx.doi.org/https://doi.org/10.1103/RevModPhys.83.1001}{\JournalTitle{\rmp},
  83, 1001}

\bibitem[{{Purcell} {et~al.}(1997){Purcell}, {Cheng}, {Dixon}, {Kinzer},
  {Kurfess}, {Leventhal}, {Saunders}, {Skibo}, {Smith}, \&
  {Tueller}}]{purcell1997}
{Purcell}, W.~R., {Cheng}, L.~X., {Dixon}, D.~D., {et~al.} 1997,
  \href{http://dx.doi.org/10.1086/304994}{\JournalTitle{\apj}, 491, 725}

\bibitem[{Richardson(1972)}]{richardson1972}
Richardson, W.~H. 1972,
  \href{http://dx.doi.org/https://doi.org/10.1364/JOSA.62.000055}{\JournalTitle{\josa},
  62, 55}

\bibitem[{{Siegert} {et~al.}(2019){Siegert}, {Crocker}, {Diehl}, {Krause},
  {Panther}, {Pleintinger}, \& {Weinberger}}]{siegert2019}
{Siegert}, T., {Crocker}, R.~M., {Diehl}, R., {et~al.} 2019,
  \href{http://dx.doi.org/10.1051/0004-6361/201833856}{\JournalTitle{\aap},
  627, A126}

\bibitem[{Siegert {et~al.}(2016)Siegert, Diehl, Khachatryan, Krause,
  Guglielmetti, Greiner, Strong, \& Zhang}]{siegert2016a}
Siegert, T., Diehl, R., Khachatryan, G., {et~al.} 2016,
  \href{http://dx.doi.org/https://doi.org/10.1051/0004-6361/201527510}{\JournalTitle{\aap},
  586, A84}

\bibitem[{Sizun {et~al.}(2006)Sizun, Cass\'{e}, \& Schanne}]{sizun2006}
Sizun, P., Cass\'{e}, M., \& Schanne, S. 2006,
  \href{http://dx.doi.org/https://doi.org/10.1103/PhysRevD.74.063514}{\JournalTitle{\prd},
  74, 063514}

\bibitem[{Skinner {et~al.}(2014)Skinner, Diehl, Zhang, Bouchet, \&
  Jean}]{skinner2014}
Skinner, G., Diehl, R., Zhang, X., Bouchet, L., \& Jean, P. 2014,
  \href{http://dx.doi.org/https://doi.org/10.22323/1.228.0054}{in 10th INTEGRAL
  Workshop: A Synergistic View of the High-Energy Sky}

\bibitem[{Skinner {et~al.}(2013)Skinner, Jean, Knödlseder, von Ballmoos,
  Leising, Milne, \& Weidenspointner}]{skinner2012}
Skinner, G., Jean, P., Knödlseder, J., {et~al.} 2013,
  \href{http://dx.doi.org/10.22323/1.176.0112}{\JournalTitle{PoS},
  INTEGRAL2012, 112}

\bibitem[{Sleator(2019)}]{sleatorthesis2019}
Sleator, C.~C. 2019, PhD thesis, University of California, Berkeley

\bibitem[{Sleator {et~al.}(2019)Sleator, Zoglauer, Lowell, Kierans, Pellegrini,
  Beechert, Boggs, Brandt, Lazar, Roberts, Siegert, \& Tomsick}]{sleator2019}
Sleator, C.~C., Zoglauer, A., Lowell, A.~W., {et~al.} 2019,
  \href{http://dx.doi.org/https://doi.org/10.1016/j.nima.2019.162643}{\JournalTitle{\nima},
  946, 162643}

\bibitem[{Snedecor \& Cochran(1991)}]{ftest}
Snedecor, G.~W., \& Cochran, W.~G. 1991, Statistical Methods, 8th edn. (Iowa
  State University Press.)

\bibitem[{Vedrenne {et~al.}(2003)Vedrenne, Roques, Sch{\"o}nfelder, Mandrou,
  Lichti, von Kienlin, Cordier, Schanne, Kn{\"o}dlseder, Skinner, Jean,
  Sanchez, Caraveo, Teegarden, von Ballmoos, Bouchet, Paul, Matteson, Boggs,
  Wunderer, Leleux, Weidenspointner, Durouchoux, Diehl, Strong, Cass{\'e},
  Clair, \& Andr{\'e}}]{vedrenne2003}
Vedrenne, G., Roques, J.~P., Sch{\"o}nfelder, V., {et~al.} 2003,
  \href{http://dx.doi.org/https://doi.org/10.1051/0004-6361:20031482}{\JournalTitle{\aap},
  411, L63}

\bibitem[{Venter {et~al.}(2015)Venter, Kopp, Harding, Gonthier, \&
  B{\"u}sching}]{venter2015}
Venter, C., Kopp, A., Harding, A.~K., Gonthier, P.~L., \& B{\"u}sching, I.
  2015,
  \href{http://dx.doi.org/https://doi.org/10.1088/0004-637X/807/2/130}{\JournalTitle{\apj},
  807, 130}

\bibitem[{{von Ballmoos} {et~al.}(1989){von Ballmoos}, {Diehl}, \&
  {Schoenfelder}}]{vonballmoos1989}
{von Ballmoos}, P., {Diehl}, R., \& {Schoenfelder}, V. 1989,
  \JournalTitle{\aap}, 221, 396

\bibitem[{Weidenspointner {et~al.}(2008)Weidenspointner, Skinner, Jean,
  Kn{\"o}dlseder, von Ballmoos, Bignami, Diehl, Strong, Cordier, Schanne, \&
  Winkler}]{weidenspointner2008}
Weidenspointner, G., Skinner, G., Jean, P., {et~al.} 2008,
  \href{http://dx.doi.org/https://doi.org/10.1038/nature06490}{\JournalTitle{\nat},
  451, 159}

\bibitem[{{Winkler} {et~al.}(2003){Winkler}, {Courvoisier}, {Di Cocco},
  {Gehrels}, {Gim{\'e}nez}, {Grebenev}, {Hermsen}, {Mas-Hesse}, {Lebrun},
  {Lund}, {Palumbo}, {Paul}, {Roques}, {Schnopper}, {Sch{\"o}nfelder},
  {Sunyaev}, {Teegarden}, {Ubertini}, {Vedrenne}, \& {Dean}}]{winkler2003}
{Winkler}, C., {Courvoisier}, T.~J.~L., {Di Cocco}, G., {et~al.} 2003,
  \href{http://dx.doi.org/10.1051/0004-6361:20031288}{\JournalTitle{\aap}, 411,
  L1}

\bibitem[{Zoglauer {et~al.}(2006)Zoglauer, Andritschke, \&
  Schopper}]{zoglauer2006}
Zoglauer, A., Andritschke, R., \& Schopper, F. 2006,
  \href{http://dx.doi.org/https://doi.org/10.1016/j.newar.2006.06.049}{\JournalTitle{\nar},
  50, 629}

\bibitem[{Zoglauer \& Kanbach(2003)}]{zoglauer2003}
Zoglauer, A., \& Kanbach, G. 2003,
  \href{http://dx.doi.org/https://doi.org/10.1117/12.461177}{\JournalTitle{Proceedings
  of SPIE}, 4851}

\bibitem[{{Zurek}(1985)}]{zurek1985}
{Zurek}, W.~H. 1985,
  \href{http://dx.doi.org/10.1086/162921}{\JournalTitle{\apj}, 289, 603}

\end{thebibliography}

\end{document}